\documentstyle[12pt,axodraw,graphicx]{article}
\begin{document}
\baselineskip 0.6cm

\def\simgt{\mathrel{\lower2.5pt\vbox{\lineskip=0pt\baselineskip=0pt
           \hbox{$>$}\hbox{$\sim$}}}}
\def\simlt{\mathrel{\lower2.5pt\vbox{\lineskip=0pt\baselineskip=0pt
           \hbox{$<$}\hbox{$\sim$}}}}
\def\x{x}

\begin{titlepage}

\begin{flushright}
FERMILAB-Pub-03/136-T \\
MIT-CTP-3375
\end{flushright}

\vskip 1.5cm

\begin{center}

{\Large \bf 
Spectrum of TeV Particles in Warped Supersymmetric Grand Unification
}

\vskip 0.6cm

{\large
Yasunori Nomura$^a$ and David R.~Smith$^b$
}

\vskip 0.3cm

$^a$ {\it Theoretical Physics Department, Fermi National Accelerator 
                Laboratory, Batavia, IL 60510} \\
$^b$ {\it Center for Theoretical Physics, Massachusetts Institute of 
                Technology, Cambridge, MA 02139}

\vskip 1.0cm

\abstract{
In warped supersymmetric grand unification, XY gauge particles 
appear near the TeV scale along with Kaluza-Klein towers of the 
standard model gauge fields. In spite of this exotic low-energy 
physics, MSSM gauge coupling unification is preserved and proton 
decay is naturally suppressed.  In this paper we study in detail 
the low-lying mass spectrum of superparticles and GUT particles 
in this theory, taking supersymmetry breaking to be localized to 
the TeV brane.  The masses of the MSSM particles, Kaluza-Klein modes,
and XY states are all determined by two parameters, one which fixes 
the strength of the supersymmetry breaking and the other which sets 
the scale of the infrared brane.  A particularly interesting result 
is that for relatively strong supersymmetry breaking, the XY gauginos 
and the lowest Kaluza-Klein excitations of the MSSM gauginos may 
both lie within reach of the LHC, providing the possibility that 
the underlying unified gauge symmetry and the enhanced $N=2$ 
supersymmetry of the theory will both be revealed.}

\end{center}
\end{titlepage}

\section{Introduction}
\label{sec:intro}

One of the most striking results of any extension of the standard 
model is the unification of gauge couplings in theories with 
a supersymmetric desert above the TeV scale.  Certain mysteries 
of the standard model, such as the stability of the Higgs 
potential and hypercharge quantization, can be elegantly addressed 
by a combination of low-energy supersymmetry and grand unification 
at high energies~\cite{Dimopoulos:1981zb}, making the supersymmetric
desert seem even more compelling.  An immediate consequence of 
this framework is the presence of superparticles at the electroweak 
scale.  In the standard paradigm these superparticles are assumed 
to be in the smallest possible supersymmetric representations: the 
standard model particles and their superpartners form multiplets 
of $N=1$ supersymmetry, and there are essentially no other fields 
at the TeV scale charged under the standard model gauge group. 
This is the basis of the minimal supersymmetric standard model (MSSM), 
which has been the main focus of phenomenological studies in 
physics beyond the standard model.

In a previous paper~\cite{Goldberger:2002pc} we studied 
an alternative to this framework with Goldberger. We constructed 
a realistic theory of grand unification in warped space, in which 
the unified gauge symmetry is broken by boundary conditions and 
the electroweak scale is generated by the warp factor.  The theory 
predicts a rich spectrum of new particles at the TeV scale; in 
addition to the usual superpartners, there are Kaluza-Klein (KK) 
towers for the standard model fields as well as their supersymmetric 
and $SU(5)$ partners.  The appearance of these particles allows 
the theory to be ``higher dimensional'' at the TeV scale. In particular, 
radiative corrections to the Higgs potential are extremely soft, 
namely exponentially shut off above the TeV scale, and we can have 
a complete understanding of the MSSM Higgs sector through the 
$U(1)_R$ symmetric structure of the theory.  We showed that the 
theory preserves the successful MSSM prediction for gauge coupling 
unification, despite the drastic departure of the matter content 
from that of the MSSM at the TeV scale.  The theory also preserves 
a number of the successes of conventional high-scale unification: 
proton decay is sufficiently suppressed and small neutrino masses 
are naturally obtained through the seesaw mechanism.  This theory 
thus naturally synthesizes two dominant approaches to physics 
beyond the standard model: Planck-cutoff and TeV-cutoff paradigms. 
Some of these features were suggested earlier in~\cite{Pomarol:2000hp}, 
and an understanding of logarithmic gauge coupling evolution in warped 
space was developed in~[\ref{Pomarol:2000hp:X}~--~\ref{Randall:2002qr:X}]. 

In this paper we study the phenomenology of the warped supersymmetric 
grand unified theory (GUT) described above. Due to the rich spectrum 
near the TeV scale, the experimental implications of the theory can 
be quite different from the conventional MSSM.  In the dual 4D 
picture of our theory supersymmetry is broken at the TeV scale 
by strong dynamics. In the 5D picture, the effects of this supersymmetry 
breaking are parameterized by operators localized on the TeV brane. 
The phenomenology of the theory then crucially depends on the form 
and size of these operators and the location of matter and Higgs 
fields in the fifth dimension.  In this paper we consider the case 
where matter fields are localized to the Planck brane, which is 
consistent with the requirements from proton decay suppression and 
gauge coupling unification.  The Higgs fields can either be localized 
to the Planck brane or propagate in the bulk, though we focus on the 
localized case in the latter part of our analysis.  The supersymmetry 
breaking operator on the TeV brane is taken to be a linear term 
for a singlet superfield in the superpotential, which was introduced 
in~\cite{Gherghetta:2000qt} and considered in the unified theory 
context in~\cite{Goldberger:2002pc,Chacko:2003tf}. This gives 
gaugino masses at tree level through an operator localized on the 
TeV brane.  At one-loop level, squarks and sleptons obtain masses 
that are insensitive to physics above the TeV scale.  These masses 
are flavor universal, so that the supersymmetric flavor problem 
is naturally solved in this setup.

An important feature of the present framework is that the masses 
for the electroweak-scale particles are determined in terms of only 
a few free parameters.  In~\cite{Goldberger:2002pc} we have shown 
that the simplest theory of warped GUTs is obtained with the TeV 
brane respecting the full $SU(5)$ symmetry and the Planck brane 
respecting only the standard model gauge symmetry.  This implies 
that the operators on the TeV brane, including the one that generates 
the gaugino masses, must respect $SU(5)$. The coefficient of this 
gaugino mass operator determines the masses for all the superparticles, 
GUT particles and KK towers, up to the overall mass scale and small 
effects from electroweak symmetry breaking. This situation is quite 
different from that in non-unified theories~\cite{Gherghetta:2000qt}, 
where we can have arbitrary values for $SU(3)_C$, $SU(2)_L$ and 
$U(1)_Y$ gaugino masses and thus have less predictive power.
An especially interesting parameter region for our theory is 
where the coefficient of the gaugino mass operator becomes large. 
In this parameter region, the spectrum appears quite different 
from that of the MSSM: the MSSM gauginos become pseudo-Dirac states 
and one of the XY gauginos becomes quite light, even lighter than 
some of the superparticles~\cite{Goldberger:2002pc}. This makes 
future experimental searches for these particles quite exciting 
and promising.

The organization of the paper is as follows.  In 
section~\ref{sec:theory} we review the theory and summarize the 
framework for our computation.  In section~\ref{sec:analysis-1} 
we calculate the masses of the superparticles and GUT particles, 
including one-loop radiative effects. We find that even 
with the present experimental bounds on the superparticle 
masses, the lightest XY gaugino is well within the reach 
of the LHC for moderately large supersymmetry breaking. 
In section~\ref{sec:analysis-2} we discuss the physical 
Higgs-boson mass and the naturalness of electroweak symmetry 
breaking. Conclusions are given in section~\ref{sec:concl}.

\section{Theory and Framework}
\label{sec:theory}

\subsection{Warped supersymmetric GUTs}
\label{subsec:WSGUT}

We begin by reviewing the warped supersymmetric grand unified theory 
of Ref.~\cite{Goldberger:2002pc}.  The theory is formulated in a warped 
5D spacetime with the extra dimension compactified on an $S^1/Z_2$ 
orbifold: $0 \leq y \leq \pi R$, where $y$ represents the coordinate 
of the extra dimension.  The metric of this space is given by
\begin{equation}
  d s^2 \equiv G_{MN} dx^M dx^N 
    = e^{-2k|y|} \eta_{\mu\nu} dx^\mu dx^\nu + dy^2.
\label{eq:metric}
\end{equation}
Here, $k$ is the AdS curvature, which is taken to be somewhat (typically 
a factor of a few) smaller than the 5D Planck scale $M_5$; the 4D Planck 
scale, $M_{\rm Pl}$, is given by $M_{\rm Pl}^2 \simeq M_5^3/k$ and we 
take $k \sim M_5 \sim M_{\rm Pl}$.  We choose $kR \sim 10$ so that the 
TeV scale is naturally generated by the AdS warp factor: $T \equiv 
k e^{-\pi kR} \sim {\rm TeV}$~\cite{Randall:1999ee}.

We consider a supersymmetric $SU(5)$ gauge theory on the above 
gravitational background.  The bulk $SU(5)$ symmetry is broken by 
boundary conditions imposed at the boundary at $y=0$. Specifically, 
the 5D gauge multiplet can be decomposed into a 4D $N=1$ vector 
superfield $V(A_\mu, \lambda)$ and a 4D $N=1$ chiral superfield 
$\Sigma(\sigma+iA_5, \lambda')$, where both $V$ and $\Sigma$ are in 
the adjoint representation of $SU(5)$. The boundary conditions for 
these fields are given by
\begin{equation}
  \pmatrix{V \cr \Sigma}(x^\mu,-y) 
  = \pmatrix{P V P^{-1} \cr -P \Sigma P^{-1}}(x^\mu,y), 
\qquad
  \pmatrix{V \cr \Sigma}(x^\mu,-y') 
  = \pmatrix{V \cr -\Sigma}(x^\mu,y'), 
\label{eq:bc-g}
\end{equation}
where $y' = y - \pi R$, and $P$ is a $5 \times 5$ matrix acting on 
gauge space: $P = {\rm diag}(+,+,+,-,-)$.  This reduces the gauge 
symmetry at the $y=0$ brane (Planck brane) to $SU(3)_C \times SU(2)_L 
\times U(1)_Y$ (321), while the 5D bulk and the $y=\pi R$ brane 
(TeV brane) respect full $SU(5)$.  After KK decomposition, the above 
boundary conditions ensure that only the 321 components of the 4D 
vector superfield, $V$, have massless modes. The typical mass scale 
for the KK towers is $T \sim {\rm TeV}$, so that the lowest KK 
excitations of the standard model gauge fields and the lightest XY 
gauge bosons both have masses of order TeV.  In fact, the KK towers 
for these gauge fields turn out to be approximately $SU(5)$ symmetric. 

The Higgs fields are introduced in the bulk as two hypermultiplets 
transforming as the fundamental representation of $SU(5)$.  Using 
the notation where a hypermultiplet is represented by two 4D $N=1$ 
chiral superfields $\Phi(\phi,\psi)$ and $\Phi^c(\phi^c,\psi^c)$ 
with the opposite gauge transformation properties, our two Higgs 
hypermultiplets can be written as $\{ H, H^c \}$ and $\{ \bar{H}, 
\bar{H}^c \}$, where $H$ and $\bar{H}^c$ transform as ${\bf 5}$ and 
$\bar{H}$ and $H^c$ transform as $\bar{\bf 5}$ under $SU(5)$. 
The boundary conditions are given by 
\begin{equation}
  \pmatrix{H \cr H^c}(x^\mu,-y) 
  = \pmatrix{-P H \cr P H^c}(x^\mu,y), 
\qquad
  \pmatrix{H \cr H^c}(x^\mu,-y') 
  = \pmatrix{H \cr -H^c}(x^\mu,y'), 
\label{eq:bc-h}
\end{equation}
for $\{ H, H^c \}$, and similarly for $\{ \bar{H}, \bar{H}^c \}$.
After KK decomposition, only the two Higgs doublets from $H$ and 
$\bar{H}$ have massless modes.  All the other KK modes, including 
those of colored Higgs fields, are massive with characteristic mass 
scale given by $T \sim {\rm TeV}$.  The masses for these modes are 
approximately $SU(5)$ symmetric as in the case of the gauge fields.
Therefore, at this stage, the mass spectrum of the theory is given as 
follows: we have a 321 vector multiplet, $V_{321}$, and the two Higgs 
doublets, $H_D$ and $\bar{H}_D$, at the massless level together with 
$SU(5)$ symmetric (and $N=2$ supersymmetric) KK towers for the gauge 
and Higgs fields with characteristic mass scale $T \sim {\rm TeV}$.

Despite the fact that XY gauge and colored Higgs fields have masses 
of order TeV, proton decay can be adequately suppressed. One way to 
achieve this is to impose baryon number, which is possible even if 
matter propagates in the bulk, and this approach allows matter 
fields to have wavefunctions spread over the extra dimension. Another, 
probably more satisfactory way is to localize matter toward the 
Planck brane -- either strictly localized as brane fields, or 
approximately localized using bulk mass parameters.  In particular, 
the latter possibility appropriately quantizes matter hypercharges 
while avoiding rapid proton decay caused by exchanges of TeV-scale 
GUT particles.

To be more explicit, the bulk matter model has the following four 
hypermultiplets for each generation: $\{ T, T^c \}({\bf 10}), 
\{T', T'^c \}({\bf 10}), \{ F, F^c \}({\bf 5}^*)$ and 
$\{ F', F'^c \}({\bf 5}^*)$, where the numbers in parentheses 
represent the transformation properties of the non-conjugated 
fields under $SU(5)$.  The boundary conditions for the matter 
fields are given similarly to the Higgs fields, Eq.~(\ref{eq:bc-h}), 
but for $\{ T, T^c \}$ and $\{T', T'^c \}$ the matrix $P$ acts on 
both $SU(5)$ fundamental indices and the overall parities under 
$y \rightarrow -y$ are taken to be opposite between $T$ and $T'$ 
multiplets and between $F$ and $F'$ multiplets (these boundary 
conditions are summarized in Table~\ref{table:bc}).
\begin{table}
\begin{center}
\begin{tabular}{|c|c|c|}
\hline
 $(p,p')$  &  gauge and Higgs fields  & 
    bulk matter fields \\ \hline
 $(+,+)$  & $V_{321}$,      $H_D$,   $\bar{H}_D$   & 
    $T_{U,E}$, $T'_Q$,     $F_D$, $F'_L$      \\ 
 $(-,-)$  & $\Sigma_{321}$, $H^c_D$, $\bar{H}^c_D$ & 
    $T^c_{U,E}$, $T'^c_Q$, $F^c_D$, $F'^c_L$  \\ 
 $(-,+)$  & $V_{X}$,        $H_T$,   $\bar{H}_T$   & 
    $T_Q$, $T'_{U,E}$,     $F_L$, $F'_D$      \\ 
 $(+,-)$  & $\Sigma_{X}$,   $H^c_T$, $\bar{H}^c_T$ & 
    $T^c_Q$, $T'^c_{U,E}$, $F^c_L$, $F'^c_D$  \\ 
\hline
\end{tabular}
\end{center}
\caption{Boundary conditions for the bulk fields under the orbifold 
 reflections.  Here, $T^{(\prime)}_{Q,U,E}$ ($F^{(\prime)}_{D,L}$) 
 are the components of $T^{(\prime)}$ ($F^{(\prime)}$) decomposed into 
 irreducible representations of the standard model gauge group. The fields 
 written in the $(p,p')$ row, $\varphi$, obey the boundary conditions 
 $\varphi(-y) = p\, \varphi(y)$ and $\varphi(-y') = p'\, \varphi(y')$.}
\label{table:bc}
\end{table}
With these boundary conditions, a complete generation, $Q, U, D, L$ 
and $E$, arises at the massless level as $T(U,E), T'(Q), F(D), F'(L)$.
The wavefunction profiles for these modes depend on the bulk 
hypermultiplet masses
\begin{equation}
  S = \int\!d^4x \int_0^{\pi R}\!\!dy \, \sqrt{-G}
    \left[ \int d^2\theta c_\Phi k \Phi \Phi^c + {\rm h.c.} \right],
\label{eq:bulk-mass}
\end{equation}
parameterized by dimensionless quantities $c_\Phi$, where $\Phi$ runs for 
$T, T', F$ and $F'$. For $c_\Phi > 1/2$, we find that the wavefunctions 
for the zero modes are strongly localized to the Planck brane as 
$e^{-(c_\Phi-1/2)k|y|}$ and that proton decay rates are sufficiently 
suppressed for $c_\Phi \simgt 1$.

\ From the 5D viewpoint, there are three local operators that can 
contribute to the low-energy 4D gauge couplings:
\begin{equation}
  S = -{1\over 4}\int\!d^4x \int_0^{\pi R}\!\!dy \, \sqrt{-G} 
    \biggl[ \frac{1}{g_B^2} F_{MN} F^{MN} 
    + 2 \delta(y) \frac{1}{\tilde{g}_{0,a}^2} {F^a}_{\mu\nu} {F^a}^{\mu\nu} 
    + 2 \delta(y - \pi R) \frac{1}{\tilde{g}_\pi^2} 
    F_{\mu\nu} F^{\mu\nu} \biggr],
\label{eq:gen-kin}
\end{equation}
where the index $a$ runs over $SU(3)_C$, $SU(2)_L$ and $U(1)_Y$ 
($a=3,2,1$, respectively). The structure of these terms are determined 
by the restricted 5D gauge symmetry, which reduces to 321 on the $y=0$ 
brane but is $SU(5)$ at all the other points in the extra dimension. 
At the fundamental scale $M_* \sim M_5$, the coefficients of these 
operators are incalculable parameters of the effective field theory. 
Therefore, one might worry that one cannot obtain any prediction 
for the low-energy gauge couplings, which in general depend on 
these unknown parameters.  This difficulty, however, is avoided 
by requiring that the theory is strongly coupled at the scale $M_*$. 
In this case the sizes of these coefficients are estimated as $1/g_B^2 
\simeq M_*/16\pi^3$ and $1/\tilde{g}_{0,a}^2 \simeq 1/\tilde{g}_\pi^2 
\simeq 1/16\pi^2$, and one finds that the low-energy prediction is 
insensitive to the parameters $\tilde{g}_{0,a}$ and $\tilde{g}_\pi$ 
evaluated at $M_*$, which encode unknown physics above the cutoff scale 
of the theory.  The prediction for the low-energy 4D gauge couplings, 
$g_a$, is then written in the form
\begin{equation}
  \frac{1}{g_a^2(T)} 
  \simeq (SU(5)\,\,\, {\rm symmetric}) 
    + \frac{1}{8 \pi^2} \Delta^a(T, k),
\label{eq:gc-low-1}
\end{equation}
where $\Delta^a(T, k)$ is the quantity whose non-universal part can be 
unambiguously computed in the effective theory.  In our theory, this 
quantity is given by
\begin{equation}
  \pmatrix{\Delta^1 \cr \Delta^2 \cr \Delta^3}(T, k)
    \simeq \pmatrix{33/5 \cr 1 \cr -3} \ln\left(\frac{k}{T}\right),
\label{eq:gc-low-2}
\end{equation}
at one-loop leading-log level, regardless of the values of the bulk 
mass parameters as long as they are larger than or equal to 1/2: 
$c_H, c_{\bar{H}}, c_T, c_{T'}, c_F, c_{F'} \geq 1/2$ (we have 
absorbed a possible $SU(5)$ symmetric piece into the first term of 
Eq.~(\ref{eq:gc-low-1})).  This is exactly the relation obtained 
in conventional 4D supersymmetric unification with the parameter 
$k$ identified with the unification scale. Therefore, we find that 
our theory preserves the successful 4D MSSM prediction for gauge 
coupling unification, despite the drastic departure of the matter 
content from the MSSM at the TeV scale~\cite{Goldberger:2002pc}.

Given that MSSM gauge coupling unification is naturally preserved 
when matter and Higgs fields have $c \geq 1/2$ and that proton 
stability is ensured when matter fields have $c \simgt 1$, it is 
natural to focus on the case with all matter fields strongly localized 
to the Planck brane.  The $c$ parameters for the Higgs multiplets 
are less constrained, but we can certainly consider the case where 
the Higgs fields are also effectively localized to the Planck brane. 
In this case the physics is well approximated by simply regarding matter 
and Higgs as brane fields, as will be done in our calculations in the 
subsequent sections. (The analysis of section~\ref{sec:analysis-1} 
assumes matter to be localized to the Planck brane but is independent 
of the Higgs profiles; parts of section~\ref{sec:analysis-2} assume 
that the Higgs fields are also localized to the Planck brane.)  
However, one should keep in mind that our analyses also apply for 
bulk matter provided the lowest KK modes are localized toward the 
Planck brane by bulk hypermultiplet masses.

Finally, we discuss the Yukawa couplings. The Yukawa couplings are 
written on the Planck brane as
\begin{equation}
  S = \int\!d^4x \int_0^{\pi R}\!\!dy \, \sqrt{-G}\, 
    2 \delta(y) \left[ \int d^2\theta \left( y_u Q U H_D 
    + y_d Q D \bar{H}_D + y_e L E \bar{H}_D \right)
    + {\rm h.c.} \right].
\label{eq:yukawa-321}
\end{equation}
Since the gauge symmetry on the $y=0$ brane is only 321, we do not have 
unwanted $SU(5)$ mass relations such as $m_s/m_d = m_\mu/m_e$. The above 
Yukawa couplings respect a $U(1)_R$ symmetry, under which the 4D 
superfields $V, \Sigma, H$ and $\bar{H}$ are neutral, $T, T^c, F, F^c, 
T', T'^c, F'$ and $F'^c$ have unit charge, and $H^c$ and $\bar{H}^c$ have 
charge $+2$.  This $U(1)_R$ forbids dangerous dimension four and five 
proton decay operators together with a potentially large supersymmetric 
mass term for the Higgs fields, thus providing a complete solution to 
the doublet-triplet splitting and proton decay problems (the $U(1)_R$ 
symmetry is broken to its $Z_2$ subgroup through supersymmetry 
breaking discussed in the next subsection, but without reintroducing 
phenomenological problems).  Small neutrino masses can be naturally 
generated by introducing right-handed neutrino fields with the Majorana 
mass terms and neutrino Yukawa couplings on the Planck brane, through 
the conventional seesaw mechanism.

\subsection{Framework for the analyses}
\label{subsec:framework}

To calculate physical quantities such as superparticle and GUT particle 
masses, we must specify how supersymmetry is broken.  We also have to 
specify a calculational scheme for computing radiative effects, which 
are quite important for the phenomenology of the theory.

Since any mass parameter on the $y=\pi R$ brane of order the fundamental 
scale appears as a TeV scale parameter in the 4D picture, it is quite 
natural to consider supersymmetry breaking on the TeV brane. 
Specifically, we introduce a supersymmetry breaking potential 
\begin{equation}
  S = \int\!d^4x \int_0^{\pi R}\!\!dy \, \sqrt{-G}\, 
    2 \delta(y-\pi R) \left[ \int d^2\theta d^2\bar{\theta} Z^\dagger Z
    + \left( \int d^2\theta \Lambda_Z^2 Z + {\rm h.c.} \right) \right],
\label{eq:Z-TeV}
\end{equation}
on the TeV brane~\cite{Gherghetta:2000qt}. Here, $Z$ is a gauge singlet 
chiral superfield, and $\Lambda_Z$ is a mass parameter of order the 
fundamental scale $M_* \sim M_5$.  The resulting supersymmetry breaking 
is transmitted to the $SU(5)$ sector through the following operator:
\begin{equation}
  S = \int\!d^4x \int_0^{\pi R}\!\!dy \, \sqrt{-G}\, 
    2 \delta(y-\pi R) \left[ \int d^2\theta \frac{\lambda}{2 M_*} 
    Z {\cal W}^\alpha {\cal W}_\alpha + {\rm h.c.} \right],
\label{eq:ZWW-TeV}
\end{equation}
where ${\cal W}^\alpha$ represents the gauge field strength superfield 
for the bulk $SU(5)$ gauge multiplet.  (The presence of both operators 
in Eqs.~(\ref{eq:Z-TeV},~\ref{eq:ZWW-TeV}) breaks the $U(1)_R$ 
symmetry discussed in the previous subsection, but this breaking 
does not reintroduce phenomenological problems such as rapid proton 
decay~\cite{Goldberger:2002pc}.)  When expanded into component fields, 
this gives the gaugino masses localized on the TeV brane
\begin{equation}
  S = \int\!d^4x \int_0^{\pi R}\!\!dy \, \sqrt{-G}\, 
    2 \delta(y-\pi R) 
    \left[ - \frac{M_\lambda}{2} \lambda^\alpha \lambda_\alpha 
    + {\rm h.c.} \right],
\label{eq:gaugino-TeV}
\end{equation}
where $M_\lambda \equiv \lambda \Lambda_Z^2/M_*$, and $\lambda^\alpha$ 
is the $SU(5)$ gaugino.  After KK decomposition, this term gives TeV scale 
masses for the 321 gauginos.  Since the full $SU(5)$ symmetry is respected 
on the TeV brane, we find that all the gaugino masses are fixed 
by the single parameter $M_\lambda$ at tree level.  The splitting 
among the three gaugino masses then arises through radiative effects.

Squarks and sleptons, which are localized to the Planck brane, obtain 
masses at one-loop level.  Because of the geometrical separation 
between supersymmetry breaking and the place where squarks and sleptons 
are located, the resulting masses are finite and calculable in the 
effective field theory.  In fact, the loop integrals are cut off at the 
scale $T$ and are insensitive to unknown UV physics.  However, 
to find the detailed structure of the mass spectrum, for example that 
coming from the splitting of the gaugino masses, we have to include 
higher order effects.

What calculational scheme should we use to compute superparticle 
masses including radiative effects?  One way of computing radiative 
corrections in truncated AdS$_5$ is to calculate them directly 
in perturbation theory, using the KK decomposed 4D theory and 
retaining all KK modes in loops.  This procedure is justified as 
long as the external momenta, $p$, are sufficiently smaller than the 
threshold for the KK towers, $T$~\cite{Goldberger:2002cz,Contino:2002kc}, 
because then the effects from unknown physics above the cutoff scale 
are suppressed by powers of $p/M'_*$. Here $M'_*$ is the cutoff scale 
on the IR brane, $M'_* \equiv M_* e^{-\pi kR}$. However, this 
procedure is not quite suitable for computing radiative corrections 
to superparticle masses, as we expect to get powers of large logarithms 
at higher loop orders, $(\alpha/4\pi)^n (\ln(k/T))^n$ for the $n$-th 
loop, which invalidate the perturbative expansions.\footnote{
For radiative corrections to the gauge couplings, the form of the 
one-loop renormalization group equations ensures that there are no 
such terms beyond one loop, if we compute corrections to $1/g^2$.}
Although general renormalization theory relates the coefficients 
of these leading logarithms, in principle allowing their summation, 
this procedure requires computation of at least the lowest loop 
diagrams containing the large logarithms: one loop diagrams for 
the gauginos and two loops for the sfermions. These calculations 
are somewhat involved, so we do not adopt this scheme for computing 
the superparticle masses.

As in 4D theories, leading-log effects can be taken into account 
using the renormalization group method.  In truncated AdS$_5$, the 
direct analog of integrating out higher momentum modes is to 
integrate out the space closest to the Planck brane.  The specific 
procedure~\cite{Lewandowski:2002rf} is summarized as follows. 
Starting from the theory where the two branes are located at $y=0$ 
and $y=\pi R$ with the AdS curvature given by $k$, we can construct 
a theory in which the two branes are located at $y=\epsilon R$ and 
$y=\pi R$ by integrating out the region $0 \leq y \leq \epsilon R$.
Then, rescaling all the mass scales of the theory as $m \rightarrow m' 
\equiv e^{-\pi \epsilon kR} m$, we obtain a theory with the AdS 
curvature given by $k'$ ($<k$), which has the same IR scale, 
$k' e^{-(\pi-\epsilon)kR} = k e^{-\pi kR} = T$, and gives the same 
low-energy predictions as the original theory.  By repeating this 
procedure, we can obtain the theory where $k'$ is sufficiently close 
to $T$ that there is no large logarithm in loop calculations. 
However, the theory obtained in this way contains a series of higher 
dimensional operators on the UV brane suppressed only by powers of 
$k' \ll k$, whose effects on predictions are $O((T/k')^n)$ for some 
power $n$.  Without knowing the coefficients of these higher dimensional 
operators, the only way of obtaining reliable predictions is to choose 
$k'$ to be somewhat larger than $T$, but then we cannot really sum 
logarithms down to the scale $T$.  Due to this limitation, we do not 
choose this ``floating cutoff'' scheme either, although one can compute 
superparticle masses in this scheme if one is satisfied with the 
precision in which one does not distinguish the values for running 
couplings at $T$ and $k' \simeq T/\varepsilon$, where $\varepsilon$ 
sets the typical size of errors in the predictions.

Instead of changing the cutoff, one can also sum up the large logarithms 
by defining the couplings of the theory using the sliding renormalization 
scale $\mu$.  Suppose we compute radiative corrections to the mass 
of a particle that is localized on the Planck brane.  If we use the 
couplings appearing in the bare Lagrangian, the resulting expression 
contains large logarithms $\ln(k/T)$.  However, these large logarithms 
can be successfully resummed if we use the couplings defined at the 
scale $\mu \sim T$, say through momentum subtraction, measured in 
terms of the 4D metric $\eta_{\mu\nu}$. This procedure effectively 
corresponds to integrating out physics above the scale $\mu$ and 
encoding it into the couplings defined at $\mu$. For this procedure 
to work, the effective theory obviously has to be valid up to $k$, 
the scale below which the large logarithms are generated. This is 
indeed the case when we compute radiative corrections to Planck-brane 
localized quantities~\cite{Goldberger:2002cz}, because on the 
Planck brane physics is essentially four dimensional up to the 
scale $k$.  This implies that large logarithms that could invalidate 
perturbative expansions, {\it i.e.} $\ln{k/T}$'s appearing in the 
superparticle mass calculation, are effectively resummed into the 
coefficients of operators located on the Planck brane. Since we did 
not lower the cutoff scale of the theory, the mass scale on this brane 
is given by $k$, and the coefficients of these operators still scale 
by powers of $k$.  Therefore, (UV insensitive) low-energy quantities, 
such as squark and slepton masses, can be reliably computed using 
the lowest dimension operators on the Planck brane with coefficients 
evaluated at the scale $\mu \sim T$. 

Calculating superparticle masses in our setup requires only the 
values of the brane localized gauge couplings, $\tilde{g}_{0,a}$ 
and $\tilde{g}_\pi$, at the 4D scale $T$.  Recall that our strong 
coupling assumption sets these couplings to be $\simeq 4\pi$ 
at the scale $M_*$ measured in terms of the 5D metric (see the 
sentences above Eq.~(\ref{eq:gc-low-1})). This corresponds 
to $1/\tilde{g}_{0,a}^2(\mu=k) \simeq 1/16\pi^2$ and 
$1/\tilde{g}_\pi^2(\mu=T) \simeq 1/16\pi^2$ in terms 
of the 4D metric, so all $\ln(k/T)$ effects can be resummed 
by running down the couplings $\tilde{g}_{0,a}$ from the 
scale $k$ down to $T$.  Since the solution to the one-loop 
renormalization group equation (RGE) takes the form
\begin{equation}
  \frac{1}{\tilde{g}_{0,a}^2(T)} 
    = \frac{1}{\tilde{g}_{0,a}^2(k)} 
    + \frac{\tilde{b}_a}{8\pi^2} \ln\left(\frac{k}{T}\right),
\label{eq:one-loop-Mp}
\end{equation}
the required couplings $\tilde{g}_{0,a}(T)$ can be obtained from 
$\tilde{b}_a$ without knowing the precise values for the initial 
couplings, $\tilde{g}_{0,a}(k)$.  

Because the non-universal part of the low-energy 4D gauge 
couplings ({\it i.e.} the differences of the three gauge 
couplings) comes only from the Planck-brane localized couplings 
$\tilde{g}_{0,a}(T)$, we know that the non-universal part 
of $\tilde{b}_a$ must be the same as that of the MSSM: 
$\tilde{b}_a - \tilde{b}_b = b^{\rm MSSM}_a - b^{\rm MSSM}_b$.
This allows us to write $\tilde{b}_a = b^{\rm MSSM}_a + \tilde{b}$, 
where $\tilde{b}$ takes a universal value for all the 321 gauge 
groups.  To determine the value of $\tilde{b}$, we focus on the 
$U(1)_Y$ component ($a=1$).  For a $U(1)$ theory, the computation 
of Ref.~\cite{Goldberger:2002cz} explicitly shows that contributions 
to $\tilde{b}_a$ are saturated by zero-mode fields for bulk scalars 
and fermions (for non-Abelian gauge fields this saturation was shown 
only for the non-universal part). We also notice that the XY gauge 
bosons and gauginos are all strongly localized to the TeV brane and 
thus do not contribute to $\tilde{b}_1$. This is sufficient to conclude 
that $\tilde{b} = 0$, because the contribution to $\tilde{b}_1$ comes 
entirely from the scalars and fermions in the bulk and on the Planck 
brane, giving $\tilde{b}_1 = b^{\rm MSSM}_1$.

Our procedure for computing superparticle masses can now be summarized 
as follows. We first integrate out physics above the scale $T$ (in the 
4D metric) by defining the couplings at the sliding renormalization 
scale $\mu \sim T$.  Using these couplings, we compute lowest order 
contributions to the superparticle masses: tree level for the gauginos 
and one loop for the sfermions. Higher loop effects are expected not 
to contain large logarithms because they are already included in the 
renormalized couplings, so that perturbation theory must work well.
The remaining uncertainty arises from possible TeV-brane operators 
suppressed by powers of $M'_* = M_* e^{-\pi kR}$. These operators, 
which are intrinsically incalculable in the effective theory, bring 
uncertainties of $O((T/M'_*)^n)$ in the predictions, where $n$ is 
some power depending on the dimension of the operator. However, 
4D Lorentz invariance implies that these corrections are at most 
of order $(k/M_*)^2 \simlt 10\%$, and we can trust our leading-order 
computations up to uncertainties of $O(10\%)$.\footnote{
Corrections to the scalar masses induced by a supersymmetry-breaking
gaugino kinetic operator on the TeV brane can be calculated using
the gaugino propagators derived in Appendix~A, by choosing 
$1/\tilde{g}_\pi^2$ appropriately.  We find that these corrections 
do not significantly affect our results below.}
Keeping this remark in mind, in the next section we compute 
the mass spectrum of the theory.

\section{Spectra for Superparticles and GUT Particles}
\label{sec:analysis-1}

In the supersymmetric limit, the massless sector of the theory consists 
of the fields of the MSSM.  The gauge multiplets propagate in the bulk, 
so the MSSM gauge bosons and gauginos are accompanied by KK towers of 
massive gauge multiplets. These massive KK levels are $N=2$ supersymmetric 
and approximately $SU(5)$ symmetric, with masses given at tree level 
by $m_n \simeq (n-1/4)\pi \, T$ for $n=1,2,\cdots$.  The lightest 
gaugino and gauge boson KK modes thus have masses $m_1 \simeq 2.4 \, T$ 
in this limit.  If they propagate in the bulk, Higgs and matter fields 
will also have KK excitations, but the analysis of this section will 
treat matter as localized to the Planck brane (or approximately localized 
with a large bulk hypermultiplet mass term).  Our results will apply 
regardless of whether or not the Higgs fields propagate in the 
bulk, however.

When supersymmetry is broken as described in section~\ref{subsec:framework},
the MSSM gauginos acquire masses at tree level and the squarks and 
sleptons obtain masses at one loop.  Supersymmetry breaking also feeds 
into the spectrum of the KK excitations in potentially crucial fashion.
In particular, the masses of the lightest XY gauginos and of the first 
KK excitations of the 321 gauginos can be pushed well below the 
$2.4 \, T$ value that applies in the supersymmetric limit, improving 
the prospects for their discovery at colliders~\cite{Goldberger:2002pc}. 
Our aim in this section is to compute the masses of the MSSM 
superparticles along with those of the lightest 321 KK states, XY 
gauge bosons, and XY gauginos, including one-loop radiative effects.

\begin{figure}
\begin{center}
\begin{picture}(470,240)(-20,-20)
 \Text(65,210)[b]{(a) $\x = 0$}
 \Line(-2,0)(130,0)
 \LongArrow(-2,0)(-2,200) \Text(-5,200)[r]{$m_n$} \Text(-7,0)[r]{0}
   \Line(-5,50)(1,50)   \Text(-7,50)[r]{$T$}
   \Line(-5,100)(1,100) \Text(-7,100)[r]{$2T$}
   \Line(-5,150)(1,150) \Text(-7,150)[r]{$3T$}
 \Text(15,-17)[b]{$A_\mu^{321}$}
   \Line(5,0)(25,0) \Vertex(15,0){2} 
   \Line(5,120)(25,120) \Vertex(15,120){2}
 \Text(40,-17)[b]{$\lambda^{321}$}
   \Line(30,0)(50,0) \Vertex(40,0){2}
   \Line(30,120)(50,120) \Vertex(35,120){2} \Vertex(45,120){2}
 \Text(65,-17)[b]{$X_\mu$}
   \Line(55,120)(75,120) \Vertex(65,120){2}
 \Text(90,-17)[b]{$\lambda^{\rm XY}$}
   \Line(80,120)(100,120) \Vertex(85,120){2} \Vertex(95,120){2}
 \Text(115,-17)[b]{$\tilde{q},\tilde{l}$}
   \Line(105,0)(125,0) \Vertex(115,0){2}
 \Text(225,210)[b]{(b) $\x \ll 1$}
 \Line(158,0)(290,0)
 \LongArrow(158,0)(158,200) \Text(155,200)[r]{$m_n$} 
   \Line(155,50)(161,50)   
   \Line(155,100)(161,100) 
   \Line(155,150)(161,150) 
 \Text(175,-17)[b]{$A_\mu^{321}$}
   \Line(165,0)(185,0) \Vertex(175,0){2} 
   \Line(165,120)(185,120) \Vertex(175,120){2}
 \Text(200,-17)[b]{$\lambda^{321}$}
   \Line(190,3)(210,3) \Vertex(200,3){2}
   \Line(190,95)(210,95) \Vertex(200,95){2} \LongArrow(200,118)(200,98)
   \Line(190,135)(210,135) \Vertex(200,135){2} \LongArrow(200,122)(200,132)
 \Text(225,-17)[b]{$X_\mu$}
   \Line(215,120)(235,120) \Vertex(225,120){2}
 \Text(250,-17)[b]{$\lambda^{\rm XY}$}
   \Line(240,95)(260,95) \Vertex(250,95){2} \LongArrow(250,118)(250,98)
   \Line(240,135)(260,135) \Vertex(250,135){2} \LongArrow(250,122)(250,132)
 \Text(275,-17)[b]{$\tilde{q},\tilde{l}$}
   \Line(265,1)(285,1) \Vertex(275,1){2}
 \DashLine(185,120)(215,120){2}
 \DashLine(210,95)(240,95){2}
 \DashLine(210,135)(240,135){2}
 \Text(385,210)[b]{(c) $\x \gg 1$}
 \Line(318,0)(450,0)
 \LongArrow(318,0)(318,200) \Text(315,200)[r]{$m_n$} 
   \Line(315,50)(321,50)   
   \Line(315,100)(321,100) 
   \Line(315,150)(321,150) 
 \Text(335,-17)[b]{$A_\mu^{321}$}
   \Line(325,0)(345,0) \Vertex(335,0){2} 
   \Line(325,120)(345,120) \Vertex(335,120){2}
 \Text(360,-17)[b]{$\lambda^{321}$}
   \Line(350,12)(370,12) \Line(350,14)(370,14) 
   \Vertex(355,12){2} \Vertex(365,14){2} 
   \LongArrow(360,1)(360,10) \LongArrow(360,117)(360,17)
   \Line(350,179)(370,179) \Vertex(355,179){2} \LongArrow(360,123)(360,176)
   \Line(350,182)(370,182) \Vertex(365,182){2} \LongArrow(360,200)(360,185)
 \Text(385,-17)[b]{$X_\mu$}
   \Line(375,120)(395,120) \Vertex(385,120){2}
 \Text(410,-17)[b]{$\lambda^{\rm XY}$}
   \Line(400,7)(420,7) \Vertex(410,7){2} \LongArrow(410,117)(410,11)
   \Line(400,179)(420,179) \Vertex(405,179){2} \LongArrow(410,123)(410,176)
   \Line(400,182)(420,182) \Vertex(415,182){2} \LongArrow(410,200)(410,185)
 \Text(435,-17)[b]{$\tilde{q},\tilde{l}$}
   \Line(425,3)(445,3) \Vertex(435,3){2}
 \DashLine(345,120)(375,120){2}
 \DashLine(370,179)(400,179){2}
 \DashLine(370,182)(400,182){2}
\end{picture}
\caption{The lowest-lying masses for the 321 gauge bosons ($A_\mu^{321}$), 
 321 gauginos ($\lambda^{321}$), XY gauge bosons ($X_\mu$), XY gauginos 
 ($\lambda^{\rm XY}$), and MSSM scalars ($\tilde{q},\,\tilde{l}$), in the 
 presence of no supersymmetry breaking (a), weak supersymmetry breaking (b), 
 and strong supersymmetry breaking (c).  Each bullet for $\lambda^{321}$ 
 and $\lambda^{\rm XY}$ represents a Majorana and Dirac degree of freedom, 
 respectively.  Dotted lines connect nearly degenerate $SU(5)$ partners, 
 and arrows indicate displacements of the mass levels relative to their 
 supersymmetric positions.  The states not shown here include the 
 standard-model quarks and leptons, the two Higgs hypermultiplets, and 
 the real scalar fields arising from the 5D gauge multiplet, which have 
 the 321 and XY quantum numbers and are degenerate with the excited 
 states of $A_\mu^{321}$ and $X_\mu$.}
\label{fig:spectrum}
\end{center}
\end{figure}
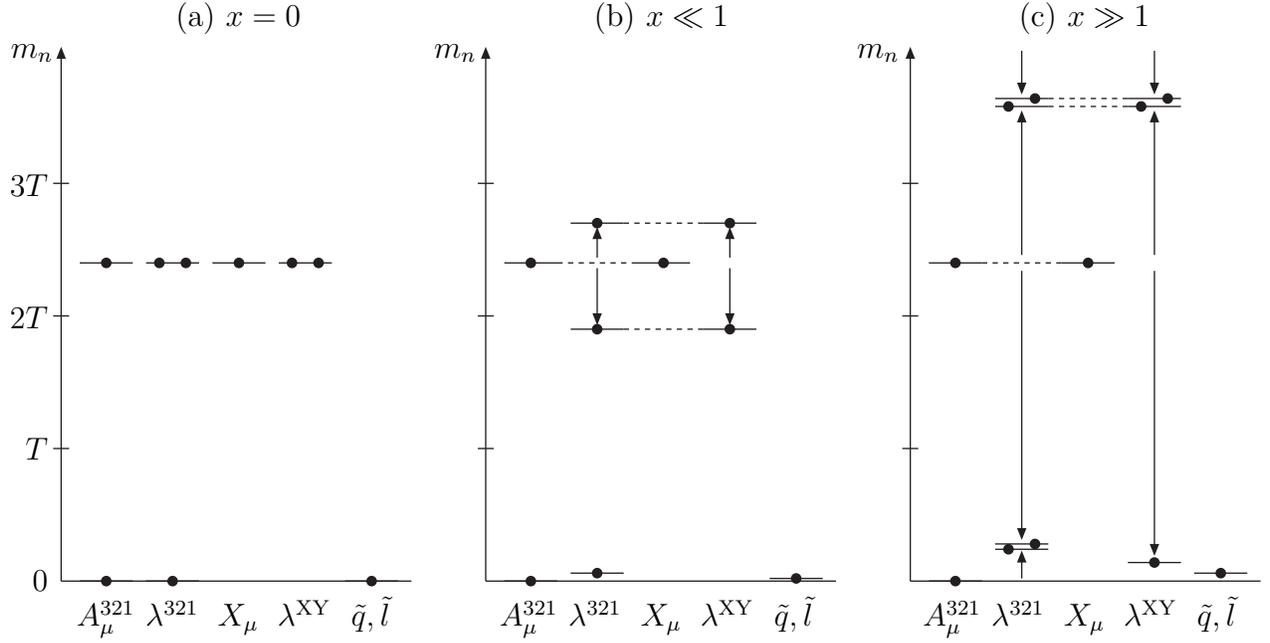
Before getting into the details of these computations, we refer the 
reader to Fig.~\ref{fig:spectrum}, where a schematic depiction of the 
effects of supersymmetry breaking on the particle spectrum is given. 
It is useful to introduce the supersymmetry-breaking parameter
\begin{equation}
  \x \equiv M_\lambda/k,
\end{equation}
which we will take to be $.01 - 10$ corresponding to relatively 
unsuppressed supersymmetry breaking on the TeV brane.  Once this 
single parameter is specified, the masses of all the particles in 
the theory are fixed up to an overall scale and small electroweak 
symmetry breaking effects.  For $\x=0$ (Fig.~\ref{fig:spectrum}a), 
the massless particles are the 321 gauge bosons $A_\mu^{321}$, the 321 
gauginos $\lambda^{321}$, and the MSSM scalars $\tilde{q}$, $\tilde{l}$ 
(and the standard-model quarks and leptons, of course).  The first KK 
level, at $m_1\simeq (3\pi/4) \,T$, includes the first $A_\mu^{321}$ 
KK mode and a degenerate pair of Majorana fermions for $\lambda^{321}$, 
which form a Dirac state.  Nearly degenerate with these are their 
$SU(5)$ partners, the lightest XY gauge bosons $X_{\mu}$ and a degenerate 
pair of Dirac XY gauginos $\lambda^{\rm XY}$.  When supersymmetry is 
broken by a small amount ($\x \ll 1$; Fig.~\ref{fig:spectrum}b), the 
$\lambda^{321}$ zero modes pick up small ($\ll T$) tree-level masses and 
the MSSM scalars pick up even smaller masses at loop level.  Meanwhile, 
the degeneracy between the $\lambda^{321}$ KK excitations is spoiled: 
one Majorana fermion's mass increases, while the other's decreases. 
The XY gaugino masses move in essentially the same way: one Dirac state 
becomes heavier while the other becomes lighter.  Thus, as long as $\x$ 
is sufficiently small, the gaugino states are still $SU(5)$ symmetric.
In the limit of very strong supersymmetry breaking ($\x \gg 1$; 
Fig.~\ref{fig:spectrum}c), the 321 gauginos once again form 
near-degenerate pairs, but now the formerly massless gaugino is 
approximately degenerate with one of the light KK excitations, with 
a mass $\simeq T/4$.  Meanwhile the other light gaugino KK excitation 
becomes nearly degenerate with one of the 321 gauginos from the second 
KK level, with mass $\simeq (5 \pi/4)\, T$.  On the other hand, the 
lightest XY gaugino has no nearly degenerate partner in this regime. 
Moreover, its mass approaches zero, rather than a finite limiting value, 
as $\x \rightarrow \infty$.  Therefore, it is possible that this XY 
gaugino is quite light.

Let us now compute the particle masses.  To calculate 
supersymmetry-breaking effects on the spectrum, we adopt the 
framework described in the previous section.  As discussed there, 
we assume that the effects of physics above the energy scale $T$ 
are encoded in local operators that reside on the Planck brane. 
The operators of interest for our purposes are the Planck-brane 
localized gauge kinetic terms.  We compute tree-level gaugino and 
gauge boson masses and one-loop scalar masses under the presence of 
these (radiatively induced) Planck-brane operators.  This requires 
knowing the numerical values of the Planck-brane, TeV-brane, and 
bulk gauge couplings at the scale $T$.  The TeV-brane coefficient 
$1/\tilde{g}_\pi^2$ is assumed to be very small at that scale, 
$\sim 1/16\pi^2$, and we neglect it entirely. The Planck-brane 
coefficients $1/\tilde{g}_{0,a}^2$ are similarly assumed to be 
negligibly small at the scale $k$, and their values at $T$ are 
obtained using the MSSM RGEs (see Eq.~(\ref{eq:one-loop-Mp}) and 
the discussion below it). Finally, the value of the bulk gauge 
coupling $g_B$ is determined from
\begin{equation}
  \frac{1}{g_a^2(T)} = \frac{\pi R}{g_B^2} 
    + \frac{1}{\tilde{g}_{0,a}^2(T)} + \frac{1}{\tilde{g}_{\pi}^2(T)},
\end{equation}
where $g_a(T)$ are the 4D gauge couplings evaluated at the scale $T$. 
These are approximated by running the experimentally measured values 
at $m_Z$ up to $\simeq 1~{\rm TeV}$ using standard model RGEs and then 
from $\simeq 1~{\rm TeV}$ to $T$ using MSSM RGEs.  Fixing in this way 
the numerical values of the various gauge couplings, we can now give 
results. 
\\

\noindent \underline{(i) 321 gauginos}
\\
The equation determining the 321 gaugino masses in the presence of 
Planck-brane and TeV-brane gauge kinetic terms is presented in 
Appendix~B.  In the limit where the TeV-brane coefficient vanishes, 
the equation becomes
\begin{equation}
  \frac{J_0\left(\frac{m_n}{k}\right)
    + \frac{g_B^2}{\tilde{g}_0^2}m_n J_1\left(\frac{m_n}{k}\right)}
  {Y_0\left(\frac{m_n}{k}\right)
    + \frac{g_B^2}{\tilde{g}_0^2}m_n Y_1\left(\frac{m_n}{k}\right)}
  = \frac{J_0\left(\frac{m_n}{T}\right)
    + g_B^2 M_\lambda J_1\left(\frac{m_n}{T}\right)}
  {Y_0\left(\frac{m_n}{T}\right)
   + g_B^2 M_\lambda Y_1\left(\frac{m_n}{T}\right)},
\label{eq:321gaugino}
\end{equation} 
where we have suppressed the 321 index. Using the values for the gauge
couplings at the scale $T$, the lowest mass solutions give the lightest 
gaugino masses at $T$, which we run down from $T$ to the gaugino masses 
themselves using MSSM RGEs. We will be interested primarily in the 
masses of the two lightest sets of gauginos.

\noindent \underline{(ii) MSSM scalars}
\\
The MSSM scalars acquire masses at one-loop level due to their gauge 
interactions.  We obtain these contributions to the scalar masses 
by computing the gaugino loop in the presence of the TeV-brane 
gaugino mass $M_{\lambda}$, and then subtracting the value of the 
same diagram in the supersymmetric limit.  The results are
\begin{eqnarray}
  m_{\tilde{q}}^2 &=& \frac{1}{2\pi^2} \left( \frac{4}{3} {\cal I}_3 
    + \frac{3}{4} {\cal I}_2 + \frac{1}{60} {\cal I}_1 \right),
\label{eq:mq2} \\
  m_{\tilde{u}}^2 &=& \frac{1}{2\pi^2} 
    \left( \frac{4}{3} {\cal I}_3 + \frac{4}{15} {\cal I}_1 \right),
\label{eq:mu2} \\
  m_{\tilde{d}}^2 &=& \frac{1}{2\pi^2} 
    \left( \frac{4}{3} {\cal I}_3 + \frac{1}{15} {\cal I}_1 \right),
\label{eq:md2} \\
  m_{\tilde{l}}^2 &=& \frac{1}{2\pi^2} 
    \left( \frac{3}{4} {\cal I}_2 + \frac{3}{20} {\cal I}_1 \right),
\label{eq:ml2} \\
  m_{\tilde{e}}^2 &=& \frac{1}{2\pi^2} 
    \left( \frac{3}{5} {\cal I}_1 \right),
\label{eq:me2} 
\end{eqnarray}
and, if the Higgs doublets are effectively localized to the Planck 
brane, $m_{h_u}^2 = m_{h_d}^2 = m_{\tilde{l}}^2$.  Here ${\cal I}_3$, 
${\cal I}_2$ and ${\cal I}_1$ are loop integrals defined as
\begin{equation}
  {\cal I}_a = \int_0^\infty\! dq\, q^3 
    \left[ f_{z,a}\Bigl(z=z'=\frac{1}{k}; q\Bigr) 
    - \left. f_{z,a}\Bigl(z=z'=\frac{1}{k}; q\Bigr)
    \right|_{M_\lambda=0} \right],
\label{eq:scint}
\end{equation}
where $a=1,2,3$ labels the gauge groups of the standard model. 
The functions $f_{z,a}(z,z';q)$ are defined in Appendix~A 
(Eq.~(\ref{eq:ApA-fz})).  As shown there, the integrand in 
Eq.~(\ref{eq:scint}) has an exponential suppression $\sim e^{-2q/T}$ 
above $T$.  This suppression at high momentum comes from the 
spatial separation between the matter fields on the Planck brane 
and the supersymmetry breaking on the TeV brane: to communicate 
the supersymmetry breaking to the MSSM scalars, the gaugino in 
the loop must propagate from the Planck brane to the TeV brane, 
where its Majorana mass term is localized, and back.

\noindent \underline{(iii) 321 KK states, XY gauge bosons, and XY gauginos}
\\
We also calculate the tree-level masses of the 321 gauge-boson KK 
excitations, XY gauge bosons, and XY gauginos.  For the 321 gauge 
bosons the equation determining the masses is 
\begin{equation}
  \frac{J_0\left(\frac{m_n}{k}\right)
    + \frac{g_B^2}{\tilde{g}_0^2}m_n J_1\left(\frac{m_n}{k}\right)}
  {Y_0\left(\frac{m_n}{k}\right)
    + \frac{g_B^2}{\tilde{g}_0^2}m_n Y_1\left(\frac{m_n}{k}\right)}
  = \frac{J_0\left(\frac{m_n}{T}\right)}
    {Y_0\left(\frac{m_n}{T}\right)},
\end{equation}
while for the XY gauge bosons we have
\begin{equation}
  \frac{J_1\left(\frac{m_n}{k}\right)}
    {Y_1\left(\frac{m_n}{k}\right)}
  = \frac{J_0\left(\frac{m_n}{T}\right)}
    {Y_0\left(\frac{m_n}{T}\right)}, 
\end{equation}
and for the XY gauginos,
\begin{equation}
  \frac{J_1\left(\frac{m_n}{k}\right)}
    {Y_1\left(\frac{m_n}{k}\right)}
  = \frac{J_0\left(\frac{m_n}{T}\right) 
    - g_B^2 M_\lambda J_1\left(\frac{m_n}{T}\right)}
  {Y_0\left(\frac{m_n}{T}\right)
   - g_B^2 M_\lambda Y_1\left(\frac{m_n}{T}\right)}.
\end{equation}
In each case we will be interested only in the lightest massive mode.

\begin{figure}[t]
  \center{\includegraphics[width=.9\textwidth]{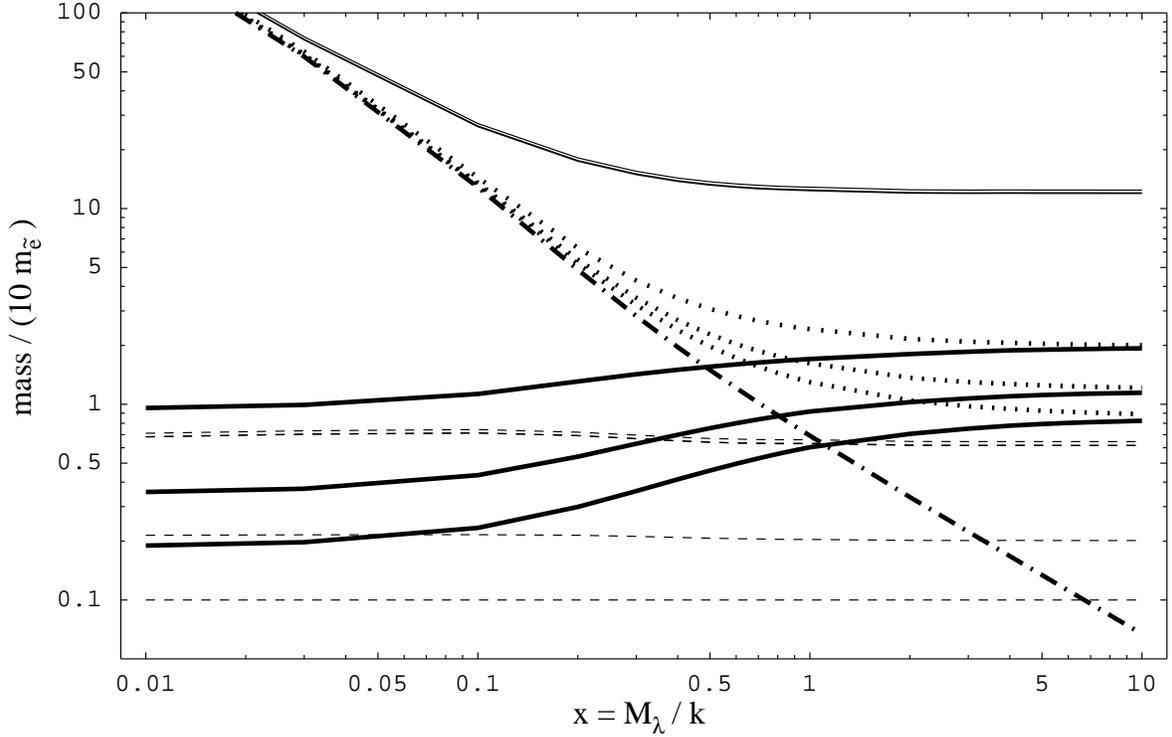}}
\caption{Masses of the MSSM scalars (dashed, with $m_{\tilde{q}}$, 
 $m_{\tilde{u}}$, and $m_{\tilde{d}}$ closely spaced and $m_{\tilde{l}}$ 
 and $m_{\tilde{e}}$ below), MSSM gauginos (thick solid), XY gauginos 
 (dot-dashed), 321 gaugino KK modes (dotted), and XY and 321 KK gauge 
 bosons (thin solid, nearly degenerate and most massive).  As explained 
 in the text, we give the masses in units of $10 \, m_{\tilde{e}}$.}
\label{fig:loglog}
\end{figure}
\begin{figure}[t]
  \center{\includegraphics[width=.9\textwidth]{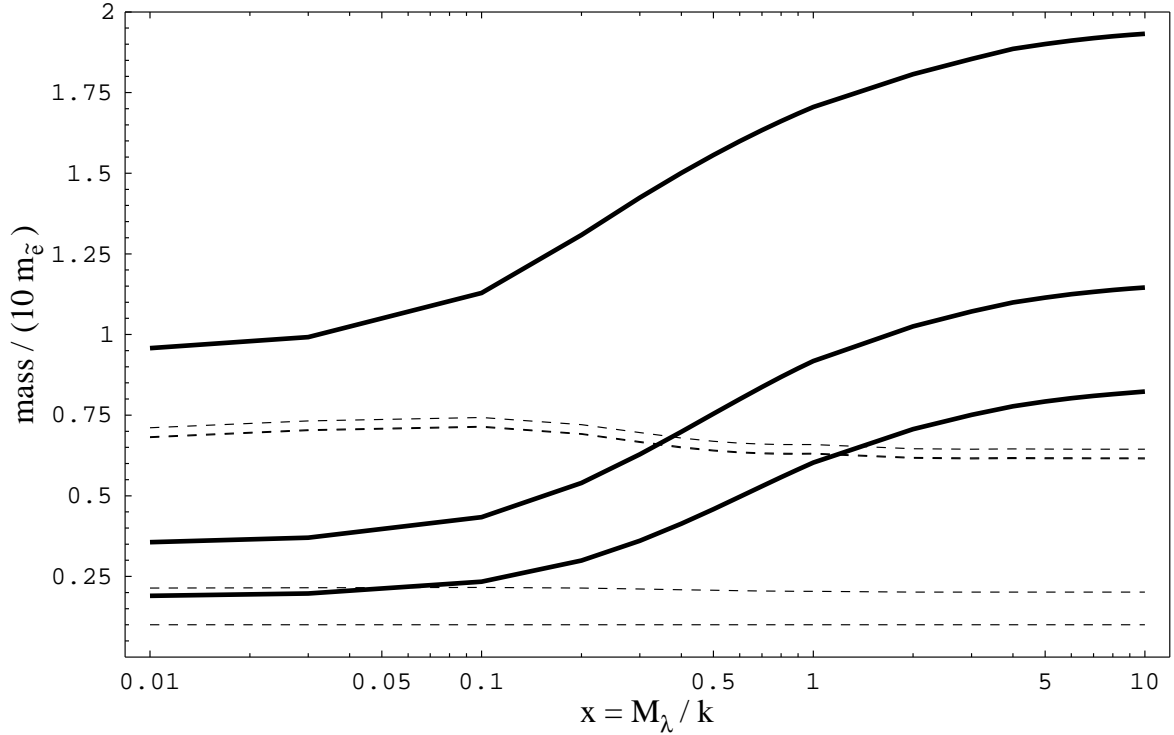}}
\caption{Masses of the MSSM scalars (dashed) and gauginos (solid).}
\label{fig:mssm}
\end{figure}
\begin{figure}[t]
  \center{\includegraphics[width=.9\textwidth]{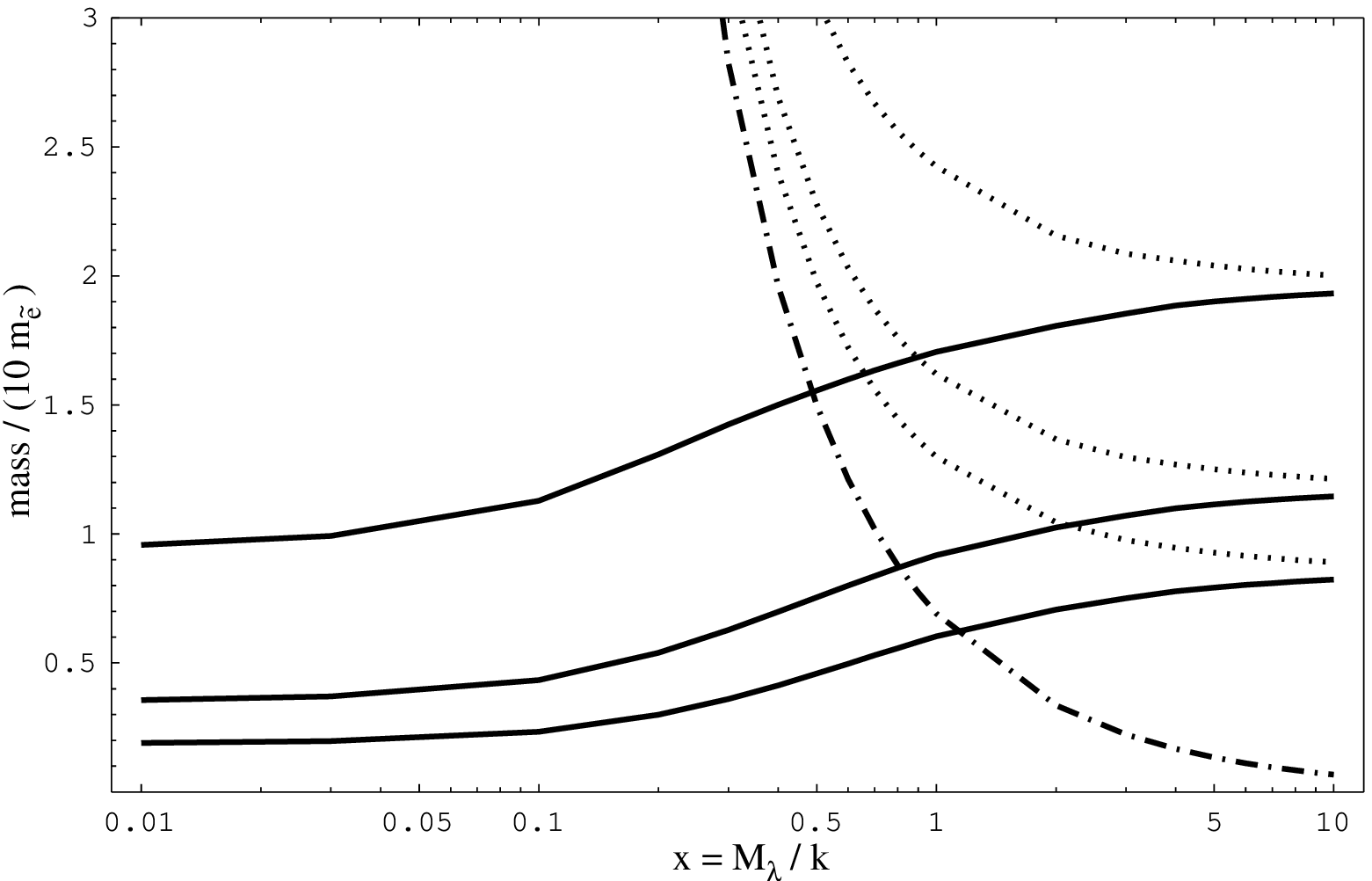}}
\caption{Masses of the MSSM gauginos (solid), XY gauginos
 (dot-dashed), and 321 gaugino KK modes (dotted).}
\label{fig:xy}
\end{figure}

The results for the various spectra are shown in 
Figs.~\ref{fig:loglog}~--~\ref{fig:xy} for $\x$ ranging from $.01$ 
to $10$.  In these figures, we have chosen to normalize the particle 
masses, $m$, in units of $10 \, m_{\tilde{e}}$ ({\it i.e.} the vertical 
axes represent $0.1\, m/m_{\tilde{e}}$), so that if the right-handed 
slepton masses are near their current experimental lower bound of 
around $100~{\rm GeV}$, the numbers labeling the vertical axes 
correspond roughly to masses in TeV units.  In Fig.~\ref{fig:loglog}, 
the masses for the MSSM particles, XY gauge bosons and gauginos, 
and KK excitations of the 321 gauge bosons and gauginos are all shown. 
The value of $T$ for a given $\x$ can be deduced from the fact that 
the XY and 321 KK gauge bosons have masses $\simeq 2.4 \,T$; $T$ ranges 
from $\simeq 100~{\rm TeV}$ on the left hand side of the plot to 
$\simeq 5~{\rm TeV}$ on the right hand side.\footnote{
For these values of $T$, constraints from the 
precision electroweak measurements ($T \simgt 
250~{\rm GeV}$~\cite{Gherghetta:2000qt,Davoudiasl:2000wi}) 
are completely negligible.}   From this plot one sees some 
overall features of the spectrum: the XY and 321 KK gauge bosons 
are quite heavy, with masses larger than $10~{\rm TeV}$ even if 
one assumes that $m_{\tilde{e}} \simeq 100~{\rm GeV}$.  On the other 
hand the masses of the 321 gaugino KK excitations decrease much more 
rapidly as $\x$ is increased, and form pseudo-Dirac states with the 
lightest gauginos at large $\x$, where the values of the masses 
plateau at $m \simeq T/4$. Finally, the XY gaugino masses are lighter 
still, and continue to decrease with increasing $\x$ even after the 
321 gaugino masses level off.  In fact, we find that for large $\x$ 
the XY gaugino masses are given by~\cite{Goldberger:2002pc}
\begin{equation}
  M_{\rm XY} \simeq \frac{2}{g_{\rm 4D}^2 \ln(k/T)\, \x} T,
\end{equation}
where $g_{\rm 4D}=g_B/\sqrt{\pi R}$ is the ``4D gauge coupling'', which 
takes a value of $O(1)$.  This fact significantly improves the discovery 
potential of the GUT particles at colliders.

Figs.~\ref{fig:mssm}~and~\ref{fig:xy} focus on the low-lying masses. 
In Fig.~\ref{fig:mssm} the masses of the MSSM scalars and gauginos 
are plotted.  From the figure we immediately see that the ratios of 
gaugino masses to scalar masses become larger for larger values for $\x$.
This is because for large $\x$ ($\x \simgt 1$) the scale of superparticle 
masses is close to the scale at which they are generated, so that the 
scalar masses are purely one-loop suppressed compared with the gaugino 
masses; on the other hand, for $\x \ll 1$, the scalar masses are enhanced 
by a logarithm between the scale of superparticle mass generation and 
the gaugino masses, $\ln(1/x)$, and become close to the gaugino masses.
Another interesting feature is that the ratios among the scalar masses 
are relatively insensitive to $\x$ for the range considered here, while 
the gaugino mass ratios change significantly as $\x$ does (their masses 
are less hierarchical for larger $\x$, a feature that may actually be 
easier to see on the log-scale plot of Fig.~\ref{fig:loglog}). 
It turns out that the gaugino masses satisfy $M_a \propto g_a^2$ for 
$\x \ll 1$ and $M_a \propto g_a$ for $\x \gg 1$~\cite{Chacko:2003tf}.
For scalar masses, we can see their rough scaling by studying the 
integrand of Eq.~(\ref{eq:scint}) evaluated at $q \sim T$, where 
the dominant contribution to the integral arises.  This suggests that
the scalar masses are approximately proportional to the square of 
the gauge couplings; in fact, we numerically find that the scalar 
masses scale roughly as the relevant Casimir times the relevant 4D 
gauge coupling squared ({\it e.g.} $(4/3) \, g_3^2$ for the squark 
masses).  Thus, the lightest among the superpartners of the standard 
model fields are the right-handed sleptons for the entire range of 
$\x$ considered.  Including effects of the Yukawa couplings, the 
lightest one will be the right-handed stau. Since the lightest 
supersymmetric particle (LSP) of the model is the gravitino $G_{3/2}$ 
with mass of order $T^2/M_{\rm Pl} \sim .01-.1~{\rm eV}$, it decays 
as $\tilde{\tau} \rightarrow \tau + G_{3/2}$ with a lifetime of 
order $8\pi T^4/m_{\tilde{e}}^5 \sim 10^{-18}-10^{-14}~{\rm sec}$.

In Fig.~\ref{fig:xy} the masses of the XY gauginos and 321 gaugino 
KK modes are shown with the MSSM gaugino masses.  We see that for
$\x \simgt 0.4$, the XY gauginos can be lighter than $2~{\rm TeV}$, 
well within the reach of the LHC.\footnote{
Although the XY particles are strongly localized to the TeV brane, 
giving exponentially small couplings to matter (and ensuring 
sufficient proton stability), their production rates are still 
unsuppressed because they have order one couplings to the 
standard model gauge fields.}
For these larger values of $\x$, the 321 gaugino KK modes also 
become light, and for much of the parameter space for which the XY 
gauginos can be discovered, these KK modes are also experimentally 
accessible.  Therefore, the experimental signatures in this parameter 
region are quite distinct: we would find two gauginos for each 321 
group, whose masses are very close for large values for $\x$, and 
one XY gaugino, which will be stable and seen as highly ionizing 
tracks at colliders~\cite{Goldberger:2002pc}.\footnote{
In fact, if the theory preserves a certain parity of the bulk 
Lagrangian called GUT parity, the XY gaugino is absolutely stable.
In the present case with matter strongly localized to the Planck 
brane, the XY gaugino is effectively stable for collider purposes 
even in the absence of GUT parity.}
This raises the exciting possibility that both the underlying 
$SU(5)$ gauge structure and $N=2$ supersymmetric structure of the 
model will be revealed by discovering these gauginos at the LHC.

\begin{figure}[t]
  \center{\includegraphics[width=.7\textwidth]{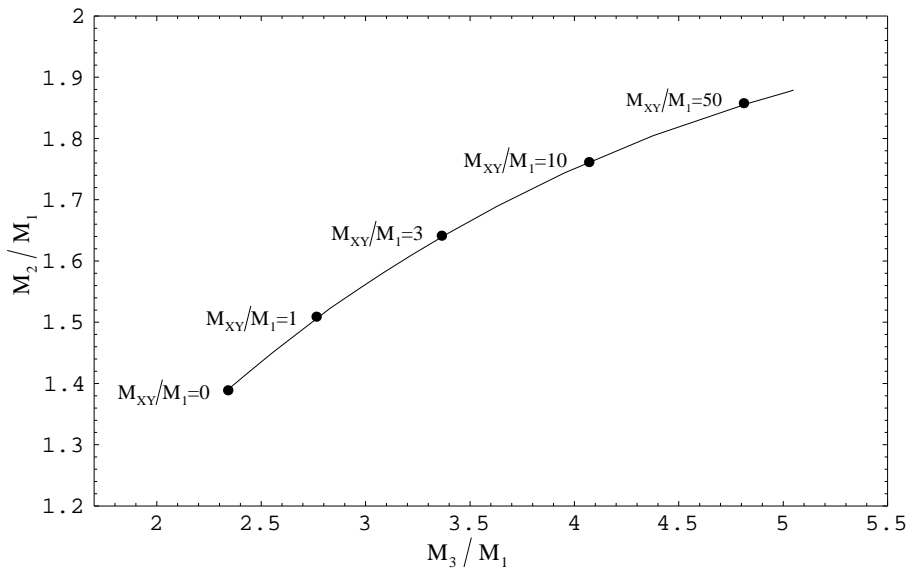}}
\caption{One gaugino mass ratio, $M_2/M_1$, given as a function of
 another, $M_3/M_1$. Each point on the line specifies a different 
 value of $\x = M_\lambda/k$, and therefore a different value of 
 $M_{\rm XY}/M_1$. The labeled points correspond to $M_{\rm XY}/M_1 
 = 0, 1, 3, 10$, and $50$.}
\label{fig:gaugino}
\end{figure}
It is interesting to point out that the three MSSM gaugino masses 
are determined by just two parameters, $T$ and $\x$, so that a single 
relation can be established among them. In particular, given the mass 
scale $T$, all the gaugino mass ratios are determined by a single 
mass ratio. This situation is depicted in Fig.~\ref{fig:gaugino}, 
where we plot $M_2/M_1$ versus $M_3/M_1$ for the fixed value 
of $m_{\tilde{e}} = 100~{\rm GeV}$. Moreover, the same two free 
parameters also determine the mass of the XY gaugino, $M_{\rm XY}$, 
so a measurement of, say, $M_3$ and $M_1$ would determine not 
only $M_2$, but also $M_{\rm XY}$.  The labeled points on the line in 
Fig.~\ref{fig:gaugino} are meant to give a sense of how $M_{\rm XY}/M_1$ 
varies as the MSSM gaugino mass ratios change. These constraints among 
the gaugino masses make this part of the spectrum especially interesting 
from an experimental standpoint.  In particular, one could predict 
the mass of the XY gaugino once the 321 gaugino masses were measured 
with sufficient accuracy.\footnote{
It is worth mentioning that the gaugino mass relations shown in 
Fig.~\ref{fig:gaugino} (and in Fig.~\ref{fig:xy} up to an overall 
mass scale) are insensitive to certain model details. In particular, 
these results apply even if matter propagates in the bulk, in which 
case squark and slepton masses can arise at tree level, allowing 
much lower values for $T$.}

We finally discuss uncertainties. As was already mentioned in 
section~\ref{subsec:framework}, higher dimensional operators on the 
TeV brane introduce an uncertainty in the superparticle masses that 
we expect to be of order $(k/M_*)^2 \simlt 10\%$.  We have used 
one-loop RGEs to evaluate the Planck-brane gauge kinetic terms 
at $T$, but errors coming from higher loops are small, especially 
because our results are not so sensitive to the relative size of 
the Planck-brane and bulk gauge couplings.  Two-loop contributions 
to the sfermion masses could be sizable, especially for $\x \ll 1$: 
in this regime the logarithmically enhanced higher-loop effects 
become larger. For $\x \simgt .01$, however, we expect that those 
are still $O(10\%)$ effects.  Based on these arguments we expect 
that our results are robust, at worst, at the $20-30\%$ level.

\section{Electroweak Symmetry Breaking}
\label{sec:analysis-2}

The naturalness of the electroweak symmetry breaking sector is an 
important issue for any extension of the standard model, and the 
original motivation for low-energy supersymmetry.  The fact that 
we have not seen physics beyond the standard model typically 
requires some degree of fine-tuning among parameters in this sector. 
In supersymmetric theories, the experimental results that are most 
relevant for naturalness considerations are (i) the non-discovery 
of superparticles, and in particular gauginos, which can easily be 
lighter than the other superparticles because their masses are 
protected by an $R$ symmetry, and (ii) the non-discovery of the 
lightest Higgs boson, which is predicted to be light in supersymmetric 
extensions of the standard model. In the MSSM, the physical mass of 
the lightest Higgs boson can be as large as $\sim 130~{\rm GeV}$, 
but for the Higgs boson to have evaded detection at LEP~II requires 
somewhat large top squark masses. These in turn generate a large 
negative contribution to the soft mass-squared parameter for the 
Higgs field, introducing some degree of fine-tuning in the 
Higgs potential.

In our theory all superparticles are heavy enough to evade experimental
bounds provided that the right-handed slepton mass is larger than 
about $100~{\rm GeV}$. This is clear from Figs.~\ref{fig:loglog} and 
\ref{fig:mssm}: the gaugino masses are significantly heavier than 
the scalars, especially for larger $\x$, because they acquire 
tree-level masses. The right-handed stau will be somewhat lighter 
than the other right-handed sleptons due to Yukawa-induced radiative
corrections (not included in our analysis), but the experimental 
bound on the stau mass is also somewhat less stringent, so we keep 
$100~{\rm GeV}$ as a representative value for the lower bound.

Now we can ask the following question: for a given value of $\x$, 
if we take $T$ large enough to evade the bound on the right-handed 
slepton mass, how heavy is the lightest Higgs boson?  Keeping all 
other parameters fixed, the Higgs mass decreases as $\x$ increases 
because the top squark masses decrease (see Fig.~\ref{fig:mssm}). 
The left-right mixing in the top squark mass matrices is dominated 
for moderately large $\tan \beta$ by a loop-generated $A_t$ term, 
\begin{equation}
  A_t = \frac{2y_t}{3\pi^2} \int_0^\infty\! dq\, q\,
    h_{z,3}\Bigl(z=z'=\frac{1}{k}; q\Bigr).
\end{equation}
Here we have included only the $SU(3)_C$ loop, $y_t$ is the 
top Yukawa coupling, and $h_{z,3}$ is the function defined 
in Appendix~A (for $SU(3)_C$).  We find that the mixing induced 
is quite small.  A no-mixing scenario requires a somewhat larger 
overall mass scale for superparticles, especially for $\x \simgt 1$, 
to obtain a sufficiently large Higgs boson mass: we estimate 
$m_{\tilde{e}} \simgt 200~{\rm GeV}$ is required for $\x \gg 1$. 
Even with these large values of superparticle masses, however, 
the fine-tuning in our theory is not as severe as one might naively 
imagine based on the squark masses alone.  This is because the scale 
of the superpartner masses is close to the the scale where they 
are generated, especially for large values of $\x$, and thus even 
if the top squark is somewhat heavy, it will not give too large 
a negative contribution to the Higgs mass-squared parameter. 
This is an interesting point: the correction to the physical 
Higgs-boson mass arises below the stop mass, and goes like 
$\ln(m_{\tilde{t}}/m_t)$, while the correction to the Higgs 
mass-squared parameter arises above the stop mass, and goes like
$\ln(M_{\rm mess}/m_{\tilde{t}})$, where $M_{\rm mess}$ is the 
scale where soft masses are generated.

An alternative way of obtaining a large Higgs boson mass is to 
introduce a singlet superfield $S$ on the Planck brane with the 
superpotential interactions $2 \delta(y) (\lambda S H_D \bar{H}_D 
+ \kappa S^3)$. This setup is also motivated as a way to naturally 
induce the supersymmetric mass term ($\mu$ term) for the Higgs 
doublets through the vacuum expectation value for the $S$ 
field.\footnote{
With these superpotential interactions, the $U(1)_R$ symmetry 
discussed in section~\ref{subsec:WSGUT} is explicitly broken to 
the $Z_{4,R}$ subgroup, under which $S$ carries a charge of $+2$.
This $Z_{4,R}$ symmetry, however, is still sufficient to forbid 
unwanted dimension four and five proton decay operators and 
a large supersymmetric mass term for the Higgs doublets.}
In this case the Higgs boson mass receives additional contribution 
at tree level thorough the coupling $\lambda$.  The size of this 
contribution depends on the value of $\lambda$, whose upper bound 
is set by Landau pole considerations.  An interesting point is 
that in our theory the 321 gauge couplings become strong at high 
energies so that the bound on $\lambda$ is significantly weaker. 
(This fact can be understood more easily in terms of the 4D dual 
picture of the theory~\cite{Goldberger:2002pc}.) This allows us 
to have the weak scale value of $\lambda$ as large as $\lambda 
\simeq 0.8$, and we obtain a large enough Higgs boson mass for 
$\tan\beta \simlt 5$ even with $m_{\tilde{e}} \simeq 100~{\rm GeV}$.

How finely tuned is electroweak symmetry breaking in our theory?
A precise discussion of naturalness requires the values of the Higgs 
mass-squared parameters, and these depend on the details of the 
Higgs sector.  Here we simply estimate their sizes for Planck-brane
localized Higgs doublets, in which case the soft masses vanish 
at tree level. The dominant radiative corrections come from 
the one-loop $SU(2)_L$ gauge contribution and the two-loop 
top-Yukawa $SU(3)_C$-gauge contribution. The former is calculated 
in the previous section (given by Eq.~(\ref{eq:ml2}) with 
$m_{h_u}^2|_{\alpha_2} = m_{h_d}^2|_{\alpha_2} = m_{\tilde{l}}^2$), 
which gives $m_{h_u}^2|_{\alpha_2} \simeq 4\, m_{\tilde{e}}^2$ 
for large $\x$. The latter we expect is similar to the flat space 
case for $\x \gg 1$, with the gluino masses in the two theories 
identified.  Referring to~\cite{Barbieri:2002sw}, we obtain 
$m_{h_u}^2|_{\alpha_3 \alpha_t} \simeq - 5\, m_{\tilde{e}}^2$, 
although the precise value is quite sensitive to the choice of the 
renormalization scale {\it etc.}, and the exact number we quote is 
not really trustworthy.  Nevertheless, it is reasonable to expect 
$m_{h_u}^2 = m_{h_u}^2|_{\alpha_2} + m_{h_u}^2|_{\alpha_3 \alpha_t} 
\simeq - c\, m_{\tilde{e}}^2$, where $c$ is a factor of a few. 
The amount of fine-tuning is given roughly by $|m_Z/m_{h_u}|^2/2$ 
for moderately large $\tan\beta$, and this is of order $10\%$ 
if we take $m_{\tilde{e}}$ at its present experimental bound. 
This is relatively mild.  The bound on the physical Higgs boson 
mass may push up the superparticle mass scale higher, and for 
smaller $\x$, the fine-tuning might change due to $\ln (1/\x)$ 
enhancements in the loop-induced scalar masses.  However, we still 
expect that the fine-tuning is not very severe.  Although this 
estimate is based on top-derived radiative electroweak symmetry 
breaking, it illustrates the required fine-tuning in our theory 
even for more general Higgs sectors.

We finally comment on the possibility of a very simple and constrained 
scenario.  Suppose that both matter and Higgs fields are localized to 
the Planck brane (either exactly or approximately, by bulk hypermultiplet 
masses), and a $\mu$ term of order the weak scale is generated on the 
Planck brane.  Then, assuming top-Yukawa driven radiative electroweak 
symmetry breaking, the entire superpartner spectrum and the parameters 
of the Higgs potential can all be calculated in terms of only three 
parameters, $\x$, $T$ and $\mu$, one of which is fixed by the observed 
value of the Higgs expectation value $v$.  In this setup, not only 
supersymmetric flavor problem, but also the supersymmetric $CP$ 
problem is solved, because the $A$ terms and the $B$ term are all 
generated radiatively through gaugino loops, and are thus all real 
in the basis where the gaugino masses are real. The sign of $\mu$ 
is determined to be negative in the standard phase convention.
A detailed study of this scenario will be interesting.

\section{Conclusions}
\label{sec:concl}

Warped supersymmetric grand unification has some remarkable features.
Most notably, it preserves MSSM-like gauge coupling unification 
even though it predicts that the exotic particles associated 
with grand unification -- XY gauge bosons, for example -- appear 
near the TeV scale.  These light exotics would normally induce 
baryon-number violating processes at disastrous rates, but 
here proton decay is naturally suppressed simply by localizing 
matter to the Planck brane.

This framework also accommodates simple, constrained possibilities 
for supersymmetry breaking.  In this paper we considered a setup 
with unsuppressed, or only mildly suppressed, supersymmetry breaking 
localized to the $SU(5)$-preserving TeV brane. In such a scenario, 
the superpartners of the standard model fields naturally acquire 
TeV-scale masses.  Taking the quark and lepton superfields to be 
localized to the Planck brane, advantageous for suppressing proton 
decay and enforcing flavor universality in the squark and slepton 
masses, one is led to a highly predictive model. In this case,
the masses of the MSSM squarks, sleptons, and gauginos, along with 
those of the XY gauge bosons and gauginos and KK excitations of 
MSSM gauge particles, are all calculable in terms of two parameters: 
$T$, the scale of the infrared brane, and $\x \equiv M_\lambda/k$, 
the ratio of the supersymmetry-breaking gaugino mass on the TeV brane 
to the curvature scale.  Calculating these masses was our main purpose.

In this setup the gauginos acquire masses at tree level and are 
heavier than the squarks and sleptons, which only acquire masses 
at loop level, and the lightest MSSM superparticles are the 
right-handed sleptons (including Yukawa effects the lightest 
one is the right-handed stau, which decays into the LSP gravitino). 
Requiring that the right-handed sleptons are heavy enough to evade 
detection at colliders sets a lower bound on $T$ for a given value 
of $\x$.  From this lower bound we infer that the XY gauge bosons 
and KK excitations of the standard model gauge bosons are too heavy 
to be detected at the LHC, with masses larger than $10~{\rm TeV}$ 
regardless of how large $\x$ becomes.  On the other hand, 
supersymmetry breaking has a dramatic impact on the masses of the 
supersymmetric partners of these gauge bosons, pushing some of the 
gaugino states to be considerably lighter. In fact, for $\x \simgt 0.4$, 
we find that the XY gauginos can have masses below $2~{\rm TeV}$, 
and for these larger values of $\x$, the KK excitations of the MSSM 
gauginos are also relatively light, and approach degeneracy with the 
MSSM gauginos themselves as $\x$ is increased.  The fact that the 
masses of these particles can be light enough to be discovered at 
the LHC -- which would reveal both the underlying unified gauge 
symmetry and the enhanced $N=2$ supersymmetry of the theory -- is 
the most important result of this paper.

Our results for the spectra of GUT particles and superparticles are 
independent of the Higgs sector, but we also briefly considered 
electroweak symmetry breaking in the particularly constrained setup 
where the Higgs doublets are localized to the Planck brane. In the 
minimal scenario, the experimental bound on the Higgs boson mass 
may require the right-handed slepton mass to be somewhat larger than 
its experimental lower bound of $100~{\rm GeV}$. On the other hand, 
a superpotential coupling between the Higgs doublets and a Planck-brane 
localized singlet $S$, motivated independently as a means for 
generating a weak-scale $\mu$ term, allows the Higgs mass to be 
raised above its lower bound quite easily. The effect on the Higgs 
mass can be stronger than in the conventional NMSSM because our theory 
has relatively strong gauge interactions at high energies, which 
drive the $S H_D \bar{H}_D$ coupling away from its Landau pole.
It will be interesting to explore electroweak symmetry breaking 
in this model in greater detail.

\section*{Acknowledgments}

We thank Nima Arkani-Hamed and Takemichi Okui for useful conversations.
The work of D.S. was supported by the U.S. Department of Energy under 
grant DE-FC02-94ER40818.

\newpage

\section*{Appendix A}

\subsection*{A.1~~~Gaugino propagators}

In this appendix we derive the propagators for the 321 gauginos in 
the presence of general brane kinetic terms and a TeV-brane localized 
Majorana mass.  The free action for the gauginos is given by
\begin{eqnarray}
  S &=& \int\!d^4x \int_0^{\pi R}\!\!dy \, 
    \Biggl\{ \frac{e^{-4ky}}{g_B^2}
    \bigg[ e^{ky}(\lambda^\dagger i\bar{\sigma}^\mu \partial_\mu \lambda 
    + \lambda' i\sigma^\mu \partial_\mu \lambda'^\dagger)
    + \lambda'(\partial_y-\frac{3k}{2})\lambda 
    + \lambda^\dagger(-\partial_y-\frac{3k}{2})\lambda'^\dagger \biggr]
\nonumber\\
  && + 2 \delta(y) \biggl[ \frac{1}{\tilde{g}_0^2}
    \lambda^\dagger i\bar{\sigma}^\mu \partial_\mu \lambda \biggr]
    + 2 \delta(y-\pi R) 
    e^{-4\pi kR} \bigg[ \frac{e^{\pi kR}}{\tilde{g}_\pi^2} 
    \lambda^\dagger i\bar{\sigma}^\mu \partial_\mu \lambda 
    - \frac{M_\lambda}{2} \lambda \lambda 
    - \frac{M_\lambda}{2} \lambda^\dagger \lambda^\dagger 
    \biggr] \Biggr\},
\label{eq:ApA-free-gaugino}
\end{eqnarray}
where $\lambda$ and $\lambda'$ are the two-component gaugino fields 
that are contained in 4D superfields $V$ and $\Sigma$, respectively.
Here, we have suppressed the index $a$ running over $SU(3)_C$, 
$SU(2)_L$ and $U(1)_Y$ for the simplicity of the notation, but it 
should be understood that the Planck-brane localized kinetic term, 
$\tilde{g}_0$, takes different values for $SU(3)_C$, $SU(2)_L$ and 
$U(1)_Y$: $\tilde{g}_0 \rightarrow \tilde{g}_{0,a}$.  The other 
parameters, $g_B$, $\tilde{g}_\pi$ and $M_\lambda$, are $SU(5)$ symmetric 
and have universal values for $SU(3)_C$, $SU(2)_L$ and $U(1)_Y$.

It is useful to define the rescaled gaugino fields $\hat{\lambda} 
\equiv e^{-2ky}\lambda$ and $\hat{\lambda}' \equiv e^{-2ky}\lambda'$.
Then, in terms of the Fourier transformed fields $\tilde{\lambda}(p,y) 
= \int d^4x \hat{\lambda}(x,y) e^{ipx}$ and $\tilde{\lambda}'(p,y) = 
\int d^4x \hat{\lambda}'(x,y) e^{ipx}$, the above action is written as
\begin{eqnarray}
  S &=& \int\!\frac{d^4p}{(2\pi)^4} \int_0^{\pi R}\!\!dy \, 
    \Biggl\{ \bigg[ \frac{e^{ky}}{g_B^2} \Bigl(
    \tilde{\lambda}^\dagger(p) \bar{\sigma}^\mu p_\mu \tilde{\lambda}(p) 
    + \tilde{\lambda}'(-p) \sigma^\mu p_\mu \tilde{\lambda}'^\dagger(-p) 
    \Bigr) 
\nonumber\\
 && + \frac{1}{g_B^2} \Bigl( 
    \tilde{\lambda}'(-p) (\partial_y+\frac{k}{2}) \tilde{\lambda}(p) 
    + \tilde{\lambda}^\dagger(p) (-\partial_y+\frac{k}{2}) 
    \tilde{\lambda}'^\dagger(-p) \Bigr) \biggr] 
    + 2 \delta(y) \biggl[ \frac{1}{\tilde{g}_0^2} \tilde{\lambda}^\dagger(p) 
    \bar{\sigma}^\mu p_\mu \tilde{\lambda}(p) \biggr]
\nonumber\\
 && + 2 \delta(y-\pi R) \bigg[ \frac{e^{\pi kR}}{\tilde{g}_\pi^2} 
    \tilde{\lambda}^\dagger(p) \bar{\sigma}^\mu p_\mu \tilde{\lambda}(p) 
    - \frac{M_\lambda}{2} \tilde{\lambda}(p) \tilde{\lambda}(-p) 
    - \frac{M_\lambda}{2} \tilde{\lambda}^\dagger(p) 
    \tilde{\lambda}^\dagger(-p) \biggr] \Biggr\}.
\label{eq:ApA-gaugino-momentum}
\end{eqnarray}
The content of this action can be divided into two parts. First, by 
examining the region near $y=0$ and $\pi R$ in the equations of motion 
derived from Eq.~(\ref{eq:ApA-gaugino-momentum}), we find the following 
conditions imposed on the fields:
\begin{eqnarray}
  && -\frac{1}{g_B^2} \tilde{\lambda}'^\dagger(-p)\biggr|_{y=\epsilon} 
    + \frac{1}{\tilde{g}_0^2} \bar{\sigma}^\mu p_\mu 
    \tilde{\lambda}(p)\biggr|_{y=0} = 0,
\label{eq:ApA-bc1} \\
  && \frac{1}{g_B^2} \sigma^\mu p_\mu 
    \tilde{\lambda}'^\dagger(-p)\biggr|_{y=\epsilon} 
    + \frac{1}{g_B^2} (\partial_y+\frac{k}{2}) 
    \tilde{\lambda}(p)\biggr|_{y=\epsilon} = 0,
\label{eq:ApA-bc2} \\
  && \frac{1}{g_B^2} \tilde{\lambda}'^\dagger(-p)\biggr|_{y=\pi R-\epsilon}
    + \frac{e^{\pi kR}}{\tilde{g}_\pi^2} \bar{\sigma}^\mu p_\mu 
    \tilde{\lambda}(p)\biggr|_{y=\pi R}
    - M_\lambda \tilde{\lambda}^\dagger(-p)\biggr|_{y=\pi R} = 0,
\label{eq:ApA-bc3} \\
  && \frac{e^{\pi kR}}{g_B^2} \sigma^\mu p_\mu 
    \tilde{\lambda}'^\dagger(-p)\biggr|_{y=\pi R-\epsilon}
    + \frac{1}{g_B^2} (\partial_y+\frac{k}{2}) 
    \tilde{\lambda}(p)\biggr|_{y=\pi R-\epsilon} = 0,
\label{eq:ApA-bc4}
\end{eqnarray}
together with the equations obtained by interchanging $\tilde{\lambda}(p) 
\leftrightarrow \tilde{\lambda}^\dagger(-p)$, $\tilde{\lambda}'(p) 
\leftrightarrow \tilde{\lambda}'^\dagger(-p)$ and $\sigma^\mu 
\leftrightarrow \bar{\sigma}^\mu$ in the above equations; here 
$\epsilon \rightarrow 0$.  These conditions provide boundary conditions 
for the corresponding propagators because the propagators are given 
by the time-ordered products of the two fields evaluated on the vacuum: 
$G_{\varphi\varphi'} = \langle 0 | T\{ \varphi\varphi' \} | 0 \rangle$.
Second, the bulk piece (the terms that do not involve delta functions) 
of Eq.~(\ref{eq:ApA-gaugino-momentum}) is written as
\begin{eqnarray}
  S_{\rm bulk} 
    &=& \frac{1}{2} \int\!\frac{d^4p}{(2\pi)^4} \int_0^{\pi R}\!\!dy \, 
  \left( \begin{array}{cc|cc} 
    \tilde{\lambda}^\dagger(p) & \tilde{\lambda}'(-p) & 
    \tilde{\lambda}(-p) & \tilde{\lambda}'^\dagger(p) 
  \end{array} \right)
\nonumber\\
  && \times \left( \begin{array}{cc|cc} 
    \frac{e^{ky}}{g_B^2} \bar{\sigma}^\mu p_\mu & 
    \frac{1}{g_B^2} (-\partial_y+\frac{k}{2}) & 0 & 0 \\ 
    \frac{1}{g_B^2} (\partial_y+\frac{k}{2}) & 
    \frac{e^{ky}}{g_B^2} \sigma^\mu p_\mu & 0 & 0 \\ \hline
    0 & 0 & \frac{e^{ky}}{g_B^2} \sigma^\mu p_\mu & 
    \frac{1}{g_B^2} (-\partial_y+\frac{k}{2}) \\
    0 & 0 & \frac{1}{g_B^2} (\partial_y+\frac{k}{2}) & 
    \frac{e^{ky}}{g_B^2} \bar{\sigma}^\mu p_\mu 
  \end{array} \right)
  \left( \begin{array}{c}
    \tilde{\lambda}(p) \\ \tilde{\lambda}'^\dagger(-p) \\ \hline
    \tilde{\lambda}^\dagger(-p) \\ \tilde{\lambda}'(p) 
  \end{array} \right).
\label{eq:ApA-gaugino-bulk}
\end{eqnarray}
This piece dictates the form of the propagators in the bulk. Defining 
the $2 \times 2$ matrix appearing in Eq.~(\ref{eq:ApA-gaugino-bulk}) 
as $M_{\rm bulk}$, the propagators 
\begin{equation}
  \hat{G} \equiv \left( \begin{array}{cc|cc} 
    \hat{G}_{\lambda \lambda^\dagger}(y,y';p) & 
    \hat{G}_{\lambda \lambda'}(y,y';p) & 
    \hat{G}_{\lambda \lambda}(y,y';p) & 
    \hat{G}_{\lambda \lambda'^\dagger}(y,y';p) \\
    \hat{G}_{\lambda'^\dagger \lambda^\dagger}(y,y';p) & 
    \hat{G}_{\lambda'^\dagger \lambda'}(y,y';p) & 
    \hat{G}_{\lambda'^\dagger \lambda}(y,y';p) & 
    \hat{G}_{\lambda'^\dagger \lambda'^\dagger}(y,y';p) \\ \hline
    \hat{G}_{\lambda^\dagger \lambda^\dagger}(y,y';p) & 
    \hat{G}_{\lambda^\dagger \lambda'}(y,y';p) & 
    \hat{G}_{\lambda^\dagger \lambda}(y,y';p) & 
    \hat{G}_{\lambda^\dagger \lambda'^\dagger}(y,y';p) \\
    \hat{G}_{\lambda' \lambda^\dagger}(y,y';p) & 
    \hat{G}_{\lambda' \lambda'}(y,y';p) & 
    \hat{G}_{\lambda' \lambda}(y,y';p) & 
    \hat{G}_{\lambda' \lambda'^\dagger}(y,y';p) 
  \end{array} \right),
\label{eq:ApA-G}
\end{equation}
are given as a solution of 
\begin{equation}
  M_{\rm bulk} \cdot \hat{G} = i \delta(y-y')\, {\bf 1},
\label{eq:ApA-prop-def}
\end{equation}
in the bulk, where ${\bf 1}$ is the $4 \times 4$ unit matrix. 
Note that $\hat{G}$ represents propagators for the rescaled fields, 
$\hat{\lambda}$ and $\hat{\lambda}'$, which are related to propagators 
$G$ for the unrescaled fields, $\lambda$ and $\lambda'$, as 
$G = e^{2k(y+y')}\hat{G}$.

Now, let us derive the bulk propagator by solving 
Eq.~(\ref{eq:ApA-prop-def}). Parameterizing the matrix 
$\hat{G}$ of Eq.~(\ref{eq:ApA-G}) as
\begin{equation}
  \hat{G} = \left( \begin{array}{cc|cc} 
    i \sigma^\mu p_\mu\, f &
    i e^{-ky} (\partial_y-\frac{k}{2})\, f' &
    i\, h &
    \frac{i\sigma^\mu p_\mu}{p^2}
      e^{-ky} (\partial_y-\frac{k}{2})\, h' \\
    i e^{-ky} (-\partial_y-\frac{k}{2})\, f &
    i \bar{\sigma}^\mu p_\mu\, f' &
    \frac{i\bar{\sigma}^\mu p_\mu}{p^2}
      e^{-ky} (-\partial_y-\frac{k}{2})\, h &
    i\, h' \\ \hline
    i\, h &
    \frac{i\bar{\sigma}^\mu p_\mu}{p^2}
      e^{-ky} (\partial_y-\frac{k}{2})\, h' &
    i \bar{\sigma}^\mu p_\mu\, f &
    i e^{-ky} (\partial_y-\frac{k}{2})\, f' \\

    \frac{i\sigma^\mu p_\mu}{p^2}
      e^{-ky} (-\partial_y-\frac{k}{2})\, h &
    i\, h' &
    i e^{-ky} (-\partial_y-\frac{k}{2})\, f &
    i \sigma^\mu p_\mu\, f'
  \end{array} \right),
\label{eq:ApA-def-fh}
\end{equation}
where $f$, $f'$, $h$ and $h'$ are the functions of $y$, $y'$ and $p$, 
we find that the functions $f$, $f'$, $h$ and $h'$ obey the following 
equations:
\begin{eqnarray}
  && \frac{1}{g_B^2} \biggl[ p^2 e^{ky} 
    + e^{-ky}(\partial_y^2 -k\partial_y -\frac{3}{4}k^2) \biggr] 
    f(y,y';p) = \delta(y-y'),
\\
  && \frac{1}{g_B^2} \biggl[ p^2 e^{ky}
    + e^{-ky}(\partial_y^2 -k\partial_y +\frac{1}{4}k^2) \biggr] 
    f'(y,y';p) = \delta(y-y'),
\\
  && \biggl[ p^2 e^{ky} 
    + e^{-ky}(\partial_y^2 -k\partial_y -\frac{3}{4}k^2) \biggr] 
    h(y,y';p) = 0,
\\
  && \biggl[ p^2 e^{ky} 
    + e^{-ky}(\partial_y^2 -k\partial_y +\frac{1}{4}k^2) \biggr] 
    h'(y,y';p) = 0.
\end{eqnarray}
Changing the variable from $y$ to $z = e^{ky}/k$, the equations become
\begin{eqnarray}
  \frac{1}{g_B^2} \biggl[ p^2 + \partial_z^2 
    -\frac{3}{4}\frac{1}{z^2} \biggr] f_z(z,z';p) &=& \delta(z-z'),
\label{eq:ApA-DE1} \\
  \frac{1}{g_B^2} \biggl[ p^2 + \partial_z^2 
    +\frac{1}{4}\frac{1}{z^2} \biggr] f'_z(z,z';p) &=& \delta(z-z'),
\label{eq:ApA-DE2} \\
  \biggl[ p^2 + \partial_z^2 -\frac{3}{4}\frac{1}{z^2} \biggr]
    h_z(z,z';p) &=& 0,
\label{eq:ApA-DE3} \\
  \biggl[ p^2 + \partial_z^2 +\frac{1}{4}\frac{1}{z^2} \biggr]
    h'_z(z,z';p) &=& 0,
\label{eq:ApA-DE4}
\end{eqnarray}
where the functions with the subscript $z$ represent the functions 
obtained by changing variables from $y$ and $y'$ to $z$ and $z'$: 
$f_z(z,z';p) \equiv f(\ln(kz)/k,\ln(kz')/k;p)$ and so on. 
Considering the region $z \neq z'$, we find that the solutions 
to Eqs.~(\ref{eq:ApA-DE1}~--~\ref{eq:ApA-DE4}) are given as 
\begin{eqnarray}
  f_z(z,z';p) &=& \sqrt{z} 
    \biggl( a_>(z')\, I_1(|p|z) + b_>(z')\, K_1(|p|z) \biggr),
\\
  f'_z(z,z';p) &=& \sqrt{z} 
    \biggl( a'_>(z')\, I_0(|p|z) + b'_>(z')\, K_0(|p|z) \biggr),
\\
  h_z(z,z';p) &=& \sqrt{z} 
    \biggl( \alpha_>(z')\, I_1(|p|z) + \beta_>(z')\, K_1(|p|z) \biggr),
\\
  h'_z(z,z';p) &=& \sqrt{z} 
    \biggl( \alpha'_>(z')\, I_0(|p|z) + \beta'_>(z')\, K_0(|p|z) \biggr),
\end{eqnarray}
for $z>z'$, where $I_n(x)$ and $K_n(x)$ are the modified Bessel 
functions of order $n$ and $|p| \equiv \sqrt{-p^2}$.  For $z<z'$ 
the functions $a_>$, $b_>$, $a'_>$, $b'_>$, $\alpha_>$, $\beta_>$, 
$\alpha'_>$ and $\beta'_>$ take different forms, which we denote 
as $a_<$, $b_<$, $a'_<$, $b'_<$, $\alpha_<$, $\beta_<$, $\alpha'_<$ 
and $\beta'_<$, respectively. 

The propagators must obey the conditions of 
Eqs.~(\ref{eq:ApA-bc1}~--~\ref{eq:ApA-bc4}). This allows us to 
solve the functions $b_>$, $b'_>$, $\beta_>$ and $\beta'_>$
in terms of $a_>$, $a'_>$, $\alpha_>$ and $\alpha'_>$; and similarly, 
$b_<$, $b'_<$, $\beta_<$ and $\beta'_<$ in terms of $a_<$, $a'_<$, 
$\alpha_<$ and $\alpha'_<$.  Further constraints on the remaining 
functions come from the continuity of the propagators at $z=z'$:
\begin{eqnarray}
  f_z(z,z';p)|_{z=z'+\epsilon} &=& f_z(z,z';p)|_{z=z'-\epsilon},
\\
  f'_z(z,z';p)|_{z=z'+\epsilon} &=& f'_z(z,z';p)|_{z=z'-\epsilon},
\\
  h_z(z,z';p)|_{z=z'+\epsilon} &=& h_z(z,z';p)|_{z=z'-\epsilon},
\\
  h'_z(z,z';p)|_{z=z'+\epsilon} &=& h'_z(z,z';p)|_{z=z'-\epsilon},
\end{eqnarray}
and the junction conditions following from 
Eqs.~(\ref{eq:ApA-DE1}~--~\ref{eq:ApA-DE4}) at $z=z'$:
\begin{eqnarray}
  \partial_z f_z(z,z';p)|_{z=z'+\epsilon} 
    - \partial_z f_z(z,z';p)|_{z=z'-\epsilon} &=& g_B^2,
\\
  \partial_z f'_z(z,z';p)|_{z=z'+\epsilon} 
    - \partial_z f'_z(z,z';p)|_{z=z'-\epsilon} &=& g_B^2,
\\
  \partial_z h_z(z,z';p)|_{z=z'+\epsilon} 
    - \partial_z h_z(z,z';p)|_{z=z'-\epsilon} &=& 0,
\\
  \partial_z h'_z(z,z';p)|_{z=z'+\epsilon} 
    - \partial_z h'_z(z,z';p)|_{z=z'-\epsilon} &=& 0.
\end{eqnarray}
These equations completely determine the functions $a_>$, $a'_>$, 
$\alpha_>$, $\alpha'_>$, $a_<$, $a'_<$, $\alpha_<$ and $\alpha'_<$.

After some algebra, we finally find that the gaugino propagators 
defined in Eq.~(\ref{eq:ApA-G}) are given by Eq.~(\ref{eq:ApA-def-fh}) 
with 
\begin{eqnarray}
  f_z(z,z';p)  &=& \frac{g_B^2 \sqrt{z_< z_>}}{(C-A)^2+B^2} 
    \biggl( I_1(|p|z_<)+C K_1(|p|z_<) \biggr) \nonumber\\
  && \times \biggl( (C-A) \Bigl\{ I_1(|p|z_>)+A K_1(|p|z_>) \Bigr\}
    - B^2 K_1(|p|z_>) \biggr),
\label{eq:ApA-fz} \\
  f'_z(z,z';p) &=& -\frac{g_B^2 \sqrt{z_< z_>}}{(C-A)^2+B^2} 
    \biggl( I_0(|p|z_<)-C K_0(|p|z_<) \biggr) \nonumber\\
  && \times \biggl( (C-A) \Bigl\{ I_0(|p|z_>)-A K_0(|p|z_>) \Bigr\}
    + B^2 K_0(|p|z_>) \biggr),
\label{eq:ApA-fpz} \\
  h_z(z,z';p)  &=& -\frac{g_B^2 |p| \sqrt{z_< z_>}}{(C-A)^2+B^2} 
    \biggl( I_1(|p|z_<)+C K_1(|p|z_<) \biggr) B 
    \biggl( I_1(|p|z_>)+C K_1(|p|z_>) \biggr),
\label{eq:ApA-hz} \\
  h'_z(z,z';p) &=& \frac{g_B^2 |p| \sqrt{z_< z_>}}{(C-A)^2+B^2} 
    \biggl( I_0(|p|z_<)-C K_0(|p|z_<) \biggr) B 
    \biggl( I_0(|p|z_>)-C K_0(|p|z_>) \biggr),
\label{eq:ApA-hpz}
\end{eqnarray}
where $|p| = \sqrt{-p^2}$ and $z_<$ ($z_>$) is the lesser 
(greater) of $z$ and $z'$; the functions $f_z$, $f'_z$, $h_z$ and $h'_z$ 
are related to $f$, $f'$, $h$ and $h'$ by $f_z(z,z';p) = f(y,y';p)$ 
{\it etc.} with $z = e^{ky}/k$.  The coefficients $A$, $B$ and $C$ 
are given by
\begin{equation}
  A = \frac{X_I X_K - Y_I Y_K}{X_K^2 + Y_K^2}, \qquad
  B = \frac{X_I Y_K + X_K Y_I}{X_K^2 + Y_K^2}, \qquad
  C = \frac{Z_I}{Z_K}.
\end{equation}
Here,
\begin{equation}
  \left\{ \begin{array}{l}
    X_I = \frac{1}{g_B^2} I_0(\frac{|p|}{T}) 
        + \frac{|p|}{\tilde{g}_\pi^2} \frac{k}{T} I_1(\frac{|p|}{T}), \\
    X_K = \frac{1}{g_B^2} K_0(\frac{|p|}{T}) 
        - \frac{|p|}{\tilde{g}_\pi^2} \frac{k}{T} K_1(\frac{|p|}{T}),
  \end{array} \right. \quad
  \left\{ \begin{array}{l}
    Y_I = M_\lambda I_1(\frac{|p|}{T}), \\
    Y_K = M_\lambda K_1(\frac{|p|}{T}),
  \end{array} \right. \quad
  \left\{ \begin{array}{l}
    Z_I = \frac{1}{g_B^2} I_0(\frac{|p|}{k}) 
        - \frac{|p|}{\tilde{g}_0^2} I_1(\frac{|p|}{k}), \\
    Z_K = \frac{1}{g_B^2} K_0(\frac{|p|}{k}) 
        + \frac{|p|}{\tilde{g}_0^2} K_1(\frac{|p|}{k}),
  \end{array} \right.
\end{equation}
where $T \equiv k e^{-\pi k R}$.

\subsection*{A.2~~~Limiting behaviors}

We here consider various limits of the gaugino propagators. We first 
consider the ``4D limit'', in which the compactification scale is 
sent to infinity and the theory reduces to the 4D MSSM. This limit 
is obtained by taking $|p|/T \rightarrow 0$ keeping $T/k$ and 
$|p|/M_\lambda$ fixed.  The function $f_z$ then becomes
\begin{equation}
  f_z(z,z';p) \rightarrow -\frac{1}{k \sqrt{z_< z_>}} 
    \frac{g_{\rm 4D}^2}{|p|^2+(g_{\rm 4D}^2 M_\lambda \frac{T}{k})^2},
\end{equation}
where $g_{\rm 4D}$ is the 4D gauge coupling given by $1/g_{\rm 4D}^2 
= \pi R/g_B^2 + 1/\tilde{g}_0^2 + 1/\tilde{g}_\pi^2$. The MSSM 
gaugino propagator, $G^{\rm 4D}_{\lambda \lambda^\dagger}(p)$, 
is obtained by multiplying the propagator in this ``4D limit'', 
$\hat{G}_{\lambda \lambda^\dagger}(y,y';p) = 
i \sigma^\mu p_\mu f(y,y';p)$, by the MSSM gaugino wavefunction 
in $\hat{\lambda}(x,y)$, $e^{3ky/2}$, and setting $y=y'=0$. 
Considering the 4D gaugino mass is given by $M_{\lambda,{\rm 4D}} 
= g_{\rm 4D}^2 M_\lambda T/k$ (see Appendix~B), this is given by
\begin{equation}
  G^{\rm 4D}_{\lambda \lambda^\dagger}(p)
  = g_{\rm 4D}^2 \frac{i \sigma^\mu p_\mu}{p^2 - M_{\lambda,{\rm 4D}}^2}, 
\end{equation}
reproducing the 4D gaugino propagator.

We next consider the propagator with the external points both on 
the Planck brane in the limit where the momentum scale $|p|$ is much 
larger than the IR scale $T$: $|p| \gg T$. We are interested in the 
difference between the gaugino propagator in the presence and absence 
of supersymmetry breaking.  For $|p| \gg T$, the relevant quantity 
$\bar{f}(p) \equiv f_z(z=z'=1/k;p) - f_z(z=z'=1/k;p)|_{M_\lambda=0}$ 
is given by
\begin{equation}
  \bar{f}(p) = \frac{2\pi M_\lambda^2}{k g_B^4}
    \frac{\left( \frac{I_1(\frac{|p|}{k})K_0(\frac{|p|}{k}) 
    + I_0(\frac{|p|}{k})K_1(\frac{|p|}{k})}
    {\frac{1}{g_B^2}K_0(\frac{|p|}{k}) 
    + \frac{|p|}{\tilde{g}_0^2}K_1(\frac{|p|}{k})} \right)^2}
    {\left( \frac{1}{g_B^2}
    + \frac{|p|}{\tilde{g}_\pi^2}\frac{k}{T} \right)
    \left( \left( \frac{1}{g_B^2}
    + \frac{|p|}{\tilde{g}_\pi^2}\frac{k}{T} \right)^2 
    + M_\lambda^2 \right)}
    \, e^{-\frac{2|p|}{T}},
\end{equation}
which shows that $\bar{f}(p)$ is exponentially suppressed 
for $p \gg T$.  This ensures the UV insensitivity for the 
scalar masses computed in section~\ref{sec:analysis-1} (see 
Eqs.~(\ref{eq:mq2}~--~\ref{eq:scint})).

\section*{Appendix B}

In this appendix we present a formula for the 321 gaugino masses, 
applicable for any value of $\x \equiv M_\lambda/k$, in the presence 
of general brane-localized kinetic terms. We start with  the action 
given in Eq.~(\ref{eq:ApA-free-gaugino}). This action gives the 
following equations of motion in the bulk:
\begin{eqnarray}
  && \frac{e^{ky}}{g_B^2} i \bar{\sigma}^\mu \partial_\mu \hat{\lambda}
    + \frac{1}{g_B^2} (-\partial_y + \frac{k}{2}) \hat{\lambda}'^\dagger 
    = 0,
\label{eq:ApB-bulk-EOM1} \\
  && \frac{e^{ky}}{g_B^2} i \sigma^\mu \partial_\mu \hat{\lambda}'^\dagger
    + \frac{1}{g_B^2} (\partial_y + \frac{k}{2}) \hat{\lambda} 
    = 0,
\label{eq:ApB-bulk-EOM2}
\end{eqnarray}
where we have presented the equations in terms of the rescaled 
gaugino fields, $\hat{\lambda} \equiv e^{-2ky}\lambda$ and 
$\hat{\lambda}' \equiv e^{-2ky}\lambda'$.  Looking for solutions 
of the form
\begin{equation}
  \hat{\lambda}(x,y) = \sum_n \lambda_n(x) f_n^{\lambda}(y),
\qquad 
  \hat{\lambda}'(x,y) = \sum_n \lambda_n(x) f_n^{\lambda'}(y),
\end{equation}
the bulk equations of motion, 
Eqs.~(\ref{eq:ApB-bulk-EOM1},~\ref{eq:ApB-bulk-EOM2}), lead to 
the following differential equation for $f_n^{\lambda}$:
\begin{equation}
  \partial_y^2 f_n^{\lambda} - k \partial_y f_n^{\lambda}
    - \frac{3k^2}{4} f_n^{\lambda} + m_n^2 e^{2ky} f_n^{\lambda} = 0.
\end{equation}
Here, $m_n$ is the 4D masses and we have used the 4D relation 
$i\bar{\sigma}^\mu \partial_\mu \lambda_n = m_n \lambda_n^\dagger$.
The solution of this equation is given by
\begin{equation}
  f_n^\lambda(y)
    = \frac{e^{\frac{k}{2}y}}{N_n} 
    \left[ J_1 \left(\frac{m_n}{k}e^{ky} \right) 
    + b_\lambda Y_1 \left(\frac{m_n}{k}e^{ky} \right) \right],
\end{equation}
where $N_n$ and $b_\lambda$ are coefficients that do not depend on $y$.

The boundary conditions for $f_n^{\lambda}$ and $f_n^{\lambda'}$ at 
$y=0$ and $\pi R$ are given by examining the equations of motion, 
Eqs.~(\ref{eq:ApB-bulk-EOM1},~\ref{eq:ApB-bulk-EOM2}):
\begin{eqnarray}
  && - f_n^{\lambda'}\biggr|_{y=\epsilon} 
    + \frac{g_B^2}{\tilde{g}_0^2} m_n f_n^{\lambda}\biggr|_{y=0} = 0,
\\
  && m_n f_n^{\lambda'}\biggr|_{y=\epsilon} 
    + (\partial_y + \frac{k}{2}) f_n^{\lambda}\biggr|_{y=\epsilon} = 0,
\\
  && f_n^{\lambda'}\biggr|_{y=\pi R-\epsilon} 
    + \frac{g_B^2 e^{\pi kR}}{\tilde{g}_\pi^2} 
      m_n f_n^{\lambda}\biggr|_{y=\pi R}
    - g_B^2 M_\lambda f_n^{\lambda}\biggr|_{y=\pi R} = 0,
\\
  && m_n e^{\pi kR} f_n^{\lambda'}\biggr|_{y=\pi R-\epsilon} 
    + (\partial_y + \frac{k}{2}) f_n^{\lambda}\biggr|_{y=\pi R-\epsilon} = 0,
\end{eqnarray}
where $\epsilon \rightarrow 0$.  Eliminating $f_n^{\lambda'}$ gives 
boundary conditions for $f_n^{\lambda}$ at $y=0$ and $\pi R$, each of 
which determines the coefficient $b_\lambda$. Equating $b_\lambda$ 
obtained from the boundary conditions at $y=0$ with that from $y=\pi R$, 
we obtain the equation that determines the 321 gaugino masses:
\begin{equation}
  \frac{J_0 \left(\frac{m_n}{k} \right) 
    + \frac{g_B^2}{\tilde{g}_0^2} m_n J_1 \left(\frac{m_n}{k} \right)}
  {Y_0 \left(\frac{m_n}{k} \right) 
    + \frac{g_B^2}{\tilde{g}_0^2} m_n Y_1 \left(\frac{m_n}{k} \right)}
  = \frac{J_0 \left(\frac{m_n}{T} \right) 
    - \frac{g_B^2}{\tilde{g}_\pi^2} \frac{k}{T} m_n 
    J_1 \left(\frac{m_n}{T} \right) 
    + g_B^2 M_\lambda J_1 \left(\frac{m_n}{T} \right)}
  {Y_0 \left(\frac{m_n}{T} \right) 
    - \frac{g_B^2}{\tilde{g}_\pi^2} \frac{k}{T} m_n 
    Y_1 \left(\frac{m_n}{T} \right) 
    + g_B^2 M_\lambda Y_1 \left(\frac{m_n}{T} \right)},
\label{eq:ApB-321-gaugino}
\end{equation}
where $T = k e^{-\pi k R}$.

\end{document}